\documentclass[reprint,amsmath,amssymb,aps,prc,superscriptaddress,showkeys]{revtex4-2}

\usepackage{graphicx}   
\usepackage{caption}
\usepackage{subcaption}
\usepackage{bm}
\usepackage{hyperref}
\usepackage{CJKutf8}
\usepackage{ragged2e}

\begin{document}

\title{Intrinsic Spatial Position Resolution of P-type Point-Contact Germanium Detector}
\thanks{This work was supported by the National Key Research and Development Program of China (Contract No. 2023YFA1607103) and the National Natural Science Foundation of China (Contracts No. 12441512, No. 11975159, and No. 11975162) provided support for this work.}

\author{Ren-Ming-Jie Li \begin{CJK}{UTF8}{gbsn}(李任明杰)\end{CJK}}
\affiliation{College of Physics, Sichuan University, Chengdu 610065, China}

\author{Shu-Kui Liu \begin{CJK}{UTF8}{gbsn}(刘书魁)\end{CJK}}
\email[Corresponding author, ]{liusk@scu.edu.cn}
\affiliation{College of Physics, Sichuan University, Chengdu 610065, China}

\author{Shin-Ted Lin \begin{CJK}{UTF8}{gbsn}(林兴德)\end{CJK}}
\email[Corresponding author, ]{stlin@scu.edu.cn}
\affiliation{College of Physics, Sichuan University, Chengdu 610065, China}

\author{Qian-Yun Li \begin{CJK}{UTF8}{bsmi}(李倩沄)\end{CJK}}
\affiliation{College of Physics, Sichuan University, Chengdu 610065, China}

\author{Li-Tao Yang \begin{CJK}{UTF8}{gbsn}(杨丽桃)\end{CJK}}
\affiliation{Key Laboratory of Particle and Radiation Imaging (Ministry of Education) and Department of Engineering Physics, Tsinghua University, Beijing 100084, China}

\author{Qian Yue \begin{CJK}{UTF8}{gbsn}(岳骞)\end{CJK}}
\affiliation{Key Laboratory of Particle and Radiation Imaging (Ministry of Education) and Department of Engineering Physics, Tsinghua University, Beijing 100084, China}

\author{Qin Wang \begin{CJK}{UTF8}{gbsn}(王琴)\end{CJK}}
\affiliation{College of Physics, Sichuan University, Chengdu 610065, China}

\author{Han-Yu Li \begin{CJK}{UTF8}{gbsn}(李含宇)\end{CJK}}
\affiliation{College of Physics, Sichuan University, Chengdu 610065, China}

\author{Xiao-Yu Peng \begin{CJK}{UTF8}{gbsn}(彭肖宇)\end{CJK}}
\affiliation{College of Physics, Sichuan University, Chengdu 610065, China}

\author{Hao-Yang Xing  \begin{CJK}{UTF8}{gbsn}(幸浩洋)\end{CJK}}
\affiliation{College of Physics, Sichuan University, Chengdu 610065, China}

\author{Jing-Jun Zhu \begin{CJK}{UTF8}{gbsn}(朱敬军)\end{CJK}}
\affiliation{College of Physics, Sichuan University, Chengdu 610065, China}

\begin{abstract}
The p-type point-contact germanium detectors have emerged as the ideal detection technology for rare-event experiments such as direct dark matter searches and neutrinoless double beta decay, and have been verified to be capable of single-site spatial position resolution. Accurately characterizing the position-dependent pulse shape responses of the detector is a crucial prerequisite for deepening background understanding and achieving background reduction. Relying on an optimized cross-scanning localization method and a full-chain physical framework, this study extracted the pulse shape responses in critical regions of the CDEX detector, quantitatively evaluated its intrinsic spatial position resolution for the first time, and ultimately achieved the position tracing of real environmental backgrounds using the constructed pulse shape database. This study completely establishes a physical analysis closed-loop for spatial position resolution, providing critical theoretical and technical support for background analysis in future ton-scale arrays.
\end{abstract}


\maketitle

\section{Introduction}\label{sec.1}

The p-type point-contact germanium (\textit{p}PCGe) detectors, benefiting from their extremely low electronic noise, ultra-low energy thresholds, and superior energy resolution, have emerged as the ideal choice for extremely rare-event experiments \cite{lukeLowCapacitanceLarge1989a, barbeauLargemassUltralowNoise2007, somaCharacterizationPerformanceGermanium2016}, such as direct dark matter searches, neutrinoless double beta decay ($0\nu\beta\beta$), and coherent elastic neutrino-nucleus scattering (CE$\nu$NS). In early explorations searching for extremely low-energy physical signals, the CoGeNT collaboration pioneered demonstrating the tremendous potential of \textit{p}PCGe detectors in the search for light dark matter; they not only achieved a sub-keV detection threshold but also advanced the in-depth study of bulk and surface event identification techniques \cite{aalsethResultsSearchLightMass2011, aalsethSearchAnnualModulation2011}. Meanwhile, the TEXONO experiment successfully lowered the detector threshold to the hundred-electron-volt level, yielding numerous leading achievements in the field of low-energy reactor neutrino physics \cite{wongSearchNeutrinoMagnetic2007, liLimitsSpinIndependentCouplings2013}. As physics goals continuously demand higher detection sensitivities, the MAJORANA and LEGEND collaborations have progressively advanced toward larger-scale detector arrays \cite{alvisSearchNeutrinolessDouble2019, arnquistFinalResultMajorana2023, acharyaFirstResultsSearch2026, legendcollaborationLEGEND1000PreconceptualDesign2021}. During this period, advanced pulse shape analysis (PSA) algorithms \cite{cooperPulseShapeAnalysis2011, alvisMultisiteEventDiscrimination2019} were developed to discriminate single-site events (SSEs, characteristic of potential rare-event signals) from multi-site events (MSEs, characteristic of gamma backgrounds), thereby fulfilling the ultra-low background requirements of experiments such as $0\nu\beta\beta$ decay.

Similarly, the China Dark Matter Experiment (CDEX) collaboration, based at the China Jinping Underground Laboratory (CJPL), utilizes \textit{p}PCGe detectors as its core detection technology, dedicating its efforts to the research of light weakly interacting massive particles (WIMPs) and $0\nu\beta\beta$ decay in $^{76}$Ge \cite{cdexcollaborationIntroductionCDEXExperiment2013, kangStatusProspectsDeep2010, chengChinaJinpingUnderground2017}. CDEX has evolved from single-module (the CDEX-1 series) \cite{liuLimitsLightWIMPs2014, zhaoSearchLowmassWIMPs2016} to array detectors (CDEX-10) \cite{cdexcollaborationLimitsLightWeakly2018, maResultsDirectDark2020}, achieving internationally leading results, and is steadily advancing toward future hundred-kilogram and ton-scale stages (CDEX-300/1T).

Empowered by advanced front-end electronics and an excellent passive shielding system, the CDEX collaboration achieved an ultra-low analysis threshold of $\sim$160~eVee (electron equivalent energy) \cite{yangLimitsLightWIMPs2018}, and demonstrated that \textit{p}PCGe detectors are capable of rough position resolution. Furthermore, to achieve background reduction, CDEX optimized the rise-time-based bulk and surface event identification algorithm \cite{liDifferentiationBulkSurface2014, yangBulkSurfaceEvent2018} to reject surface events deposited in the dead layer. These events are highly susceptible to electron-hole pair recombination, resulting in incomplete charge collection and extremely slow rise times \cite{aguayoCharacteristicsSignalsOriginating2013}. With the improved performance of the electronics response system, a class of anomalous bulk events exhibiting a faster rising trend was observed within the sensitive volume, which was verified to originate from the bottom region near the P$^+$ electrode \cite{liIdentificationAnomalousFast2022a}. Hole carriers excited in this region can be rapidly collected by the point-contact without traversing the central weak-field region of the detector, leading to shorter pulse rise times compared to typical bulk events.

Benefiting from the unique geometric structure and internal electric field distribution of the \textit{p}PCGe detector, further investigation in this study demonstrates that the pulse shapes of the CDEX-1B detector theoretically still exhibit differences at exact positions across different regions, indicating a finer single-site spatial position resolution. Accurately understanding the position-dependent pulse shape response characteristics inside the detector and evaluating its resolution is a crucial prerequisite for deepening background understanding and achieving background reduction.

Addressing these pressing physical requirements, this paper focuses on the CDEX-1B detector and proposes a cross-scanning localization method using a collimated radiation source from two linearly independent directions to achieve waveform sampling at specific spatial positions. This method is specifically optimized for the point-contact geometry and single-channel signal collection characteristics. Simultaneously, following the same logic, we constructed a full-chain analysis workflow via Monte-Carlo and pulse shape simulations to reconstruct the experimental scanning process, ultimately characterizing the pulse shape responses at different positions within the detector successfully. Based on this simulation framework, we first quantitatively explored the intrinsic spatial position resolution of the \textit{p}PCGe detector, presenting resolution results across multiple energy regions of interest. Furthermore, utilizing a pulse shape database constructed through a combination of simulation and experiment, and following the localization and validation of the collimated source data events, we ultimately performed position tracing on environmental background events collected during the experimental commissioning phase. The analysis results aligned with the actual conditions. The successful preliminary application of this analysis technique foresees that it will not only provide a powerful tool for physical event selection and background subtraction in future complex radiation background fields, but also offer crucial scientific guidance for understanding background sources in rare-event experiments and implementing background control in detector designs, ultimately facilitating the enhancement of detection sensitivity.

The remainder of this paper is organized as follows: Sec.~\ref{sec.2} introduces the physical mechanisms of the signal response in \textit{p}PCGe detectors; Sec.~\ref{sec.3} outlines the characterization method and validation of the spatial position responses within the detector; Sec.~\ref{sec.4} describes the experimental characterization of its spatial position responses; Sec.~\ref{sec.5} investigates the intrinsic spatial position resolution of the detector; Sec.~\ref{sec.6} presents the spatial position tracing of background events and the result analysis; and finally, a summary and outlook are provided in the last section.

\section{Physical mechanisms of the signal response in \textit{p}PCGe detectors}\label{sec.2}

\begin{figure}[!htb]
    \centering
    \includegraphics[width=0.5\linewidth]{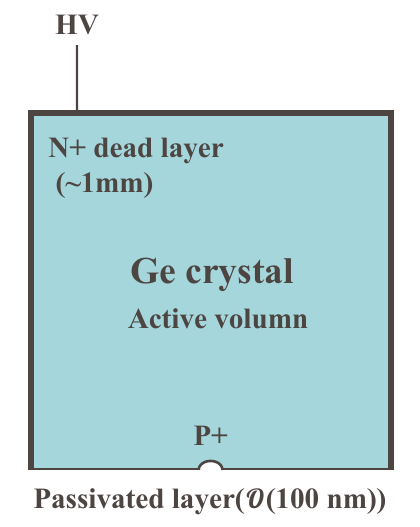}
    \caption{Simplified schematic cross-section of a typical \textit{p}PCGe detector.}
    \label{fig:1}
\end{figure}

The excellent low-noise detection performance and pulse shape discrimination capability of the \textit{p}PCGe detector essentially originate from its unique electrode geometric design. This study discusses the CDEX-1B detector as a physical model, the core of which is a cylindrical p-type high-purity germanium crystal with a mass of 939~g (illustrated in Fig.~\ref{fig:1}). The crystal measures 62.1~mm in diameter and 62.3~mm in height. The outer surface, excluding the bottom face, serves as the N$^+$ contact layer fabricated via lithium diffusion (with a dead layer thickness of $0.88\pm0.12$~mm), while the P$^+$ point-contact for signal readout shrinks to the center of the bottom face ($\mathcal{O}$(1~mm)). These specific geometric boundary conditions dictate the distortion characteristics of the internal electric field and weighting potential distribution within the crystal, which subsequently govern the dynamic behavior of the charge carriers.

\subsection{Carrier dynamics and signal formation}\label{sec.2.1}

The signal formation mechanism in germanium detectors is governed by the Shockley-Ramo theorem \cite{shockleyCurrentsConductorsInduced1938, ramoCurrentsInducedElectron1939, heReviewShockleyRamo2001}. The interaction between incident charged particles and the detector medium generates clouds of charge carriers (electron-hole pairs). Driven by the bias electric field, these carriers drift toward their respective electrodes, inducing mirror charges on the readout electrode. The transient characteristics of the induced signal are jointly determined by the carrier dynamics and the detector geometry. For a deposited charge $q$, the time-dependent evolution of the induced charge $Q(t)$ and induced current $I(t)$ can be respectively expressed as:
\begin{equation}
    Q(t) = q[\phi(\bm{r_h}(t)) - \phi(\bm{r_e}(t))],
    \label{eq:1}
\end{equation}
\begin{equation}
    I(t) = q[\bm{v_h}(\bm{r_h}(t)) \cdot \bm{E_w}(\bm{r_h}(t)) - \bm{v_e}(\bm{r_e}(t)) \cdot \bm{E_w}(\bm{r_e}(t))],
    \label{eq:2}
\end{equation}
where $\phi(\bm{r}(t))$ and $\bm{E_w}(\bm{r}(t))$ denote the weighting potential and weighting field at the carrier position $\bm{r}(t)$, respectively. These quantities are strictly determined by the detector geometry and are calculated by setting the readout electrode (the point-contact) to unit potential, setting all other electrodes (N$^+$ electrode) to zero potential, and assuming zero space charge. The term $\bm{v}(\bm{r}(t))$ represents the instantaneous drift velocity of the carriers, which is dominated by the distribution of the actual physical electric field inside the detector.

\subsection{Correlation between pulse shape characteristics and spatial position}\label{sec.2.2}

\begin{figure}[!htb]
    \centering
    \includegraphics[width=0.99\linewidth]{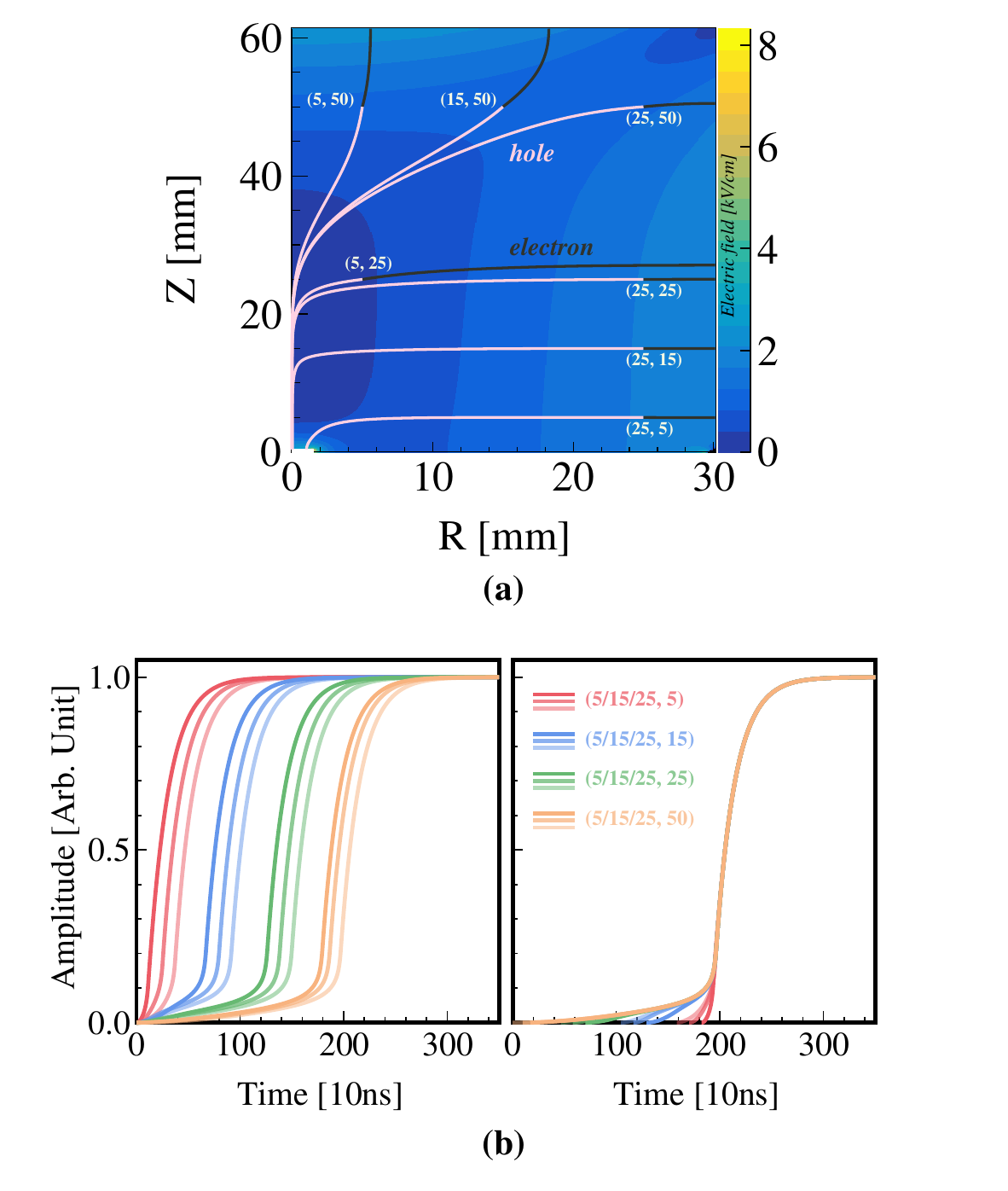}
    \caption{(a) Distribution of the weighting potential in the CDEX-1B detector. The black and pink lines represent the drift trajectories of electrons and holes, respectively. (b) Pulse signals corresponding to the different positions indicated in (a); in the right subpanel, the signals are aligned at the time point corresponding to 50\% of their maximum amplitude.}
    \label{fig:2}
\end{figure}

Due to the unique geometry of the \textit{p}PCGe detector, its internal weighting potential $\phi$ exhibits significant spatial non-uniformity. Specifically, the weighting potential remains negligible throughout the majority of the detector's sensitive volume and rises sharply to unit potential only in the immediate hemispherical vicinity of the point-contact. Fig.~\ref{fig:2}(a) illustrates the numerically simulated weighting potential distribution of the CDEX-1B detector, which profoundly dominates the evolution process of the signal:
\begin{itemize}
    \item[\textbf{[1]}] \textbf{Hole-dominated signals:} In p-type detectors, holes drift toward the central point-contact, while electrons drift toward the outer surface. Since the weighting potential in the vast peripheral region is extremely low, the contribution of electron drift to the net induced charge is negligible. Consequently, the output signal is primarily determined by the collection process of the holes.
    \item[\textbf{[2]}] \textbf{Drift phase characteristics:} When holes drift through the bulk region where $\phi \approx 0$, they induce minimal charge on the electrode. This results in a distinctive, prolonged flat baseline or an extremely slow rising component in the initial phase of the pulse shape.
    \item[\textbf{[3]}] \textbf{Fast rising edge:} Significant charge induction is stimulated only when holes enter the high weighting potential region near the point-contact. This leads to a sudden and steep rising edge at the end of the drift path.
\end{itemize}

Based on the aforementioned physical mechanisms, the pulse shape is strictly dependent on the spatial position of the initial energy deposition and the length of the carrier drift path. Events originating far from the electrode exhibit extended drift times and are referred to as bulk events (BEs), located in the $Z > 15$~mm region in Fig.~\ref{fig:2}(a); in contrast, events near the electrode lack this slow drift phase, presenting an immediate fast rising edge, and are thus termed fast bulk events (FBEs), located in $Z < 15$~mm region. Their pulses are shown in Fig.~\ref{fig:2}(b). Notably, due to the flat distribution of the weighting potential in the bulk region, pulse shapes generated in this region exhibit a high degree of homogenization (their shapes tend to be uniform); in contrast, pulse shapes for events in the fast bulk region are highly sensitive to positional changes, demonstrating superior single-site spatial position resolution.

\section{Characterization method and validation of the spatial position response within the detector}\label{sec.3}

As demonstrated in Sec.~\ref{sec.2.2}, benefiting from the unique geometric structure and internal electric field distribution of the \textit{p}PCGe detector, the pulse shapes of the CDEX-1B detector still exhibit resolvable differences in the fast bulk region and its adjacent vicinity. Accurately understanding the position-dependent pulse shape response characteristics inside the detector provides crucial scientific guidance for future background understanding and background reduction. To precisely characterize these position responses experimentally, a spatial scanning technique based on pulse shape comparison, optimized for \textit{p}PCGe detectors, has been developed.

\subsection{Optimized pulse shape comparison spatial scanning technique}\label{sec.3.1}

The pulse shape comparison scan (PSCS) technique \cite{crespiNovelTechniqueCharacterization2008a} relies on data acquired using two collimated $\gamma$-ray sources incident from specific directions (typically orthogonal) to achieve waveform sampling at specific positions within the detector. Theoretically, for position-sensitive detectors, when evaluating the two separately acquired datasets (denoted as the horizontal dataset $H$ and the vertical dataset $V$), consistency between the recorded pulse shapes is expected only when the interactions occur at the geometric intersection of the two collimated beams. Consequently, by comparing the similarity of the pulse shapes between the two datasets, highly matched events can be filtered out, thereby characterizing the pulse shape response at that specific intersection point.

\begin{figure}[!htb]
    \centering
    \includegraphics[width=0.75\linewidth]{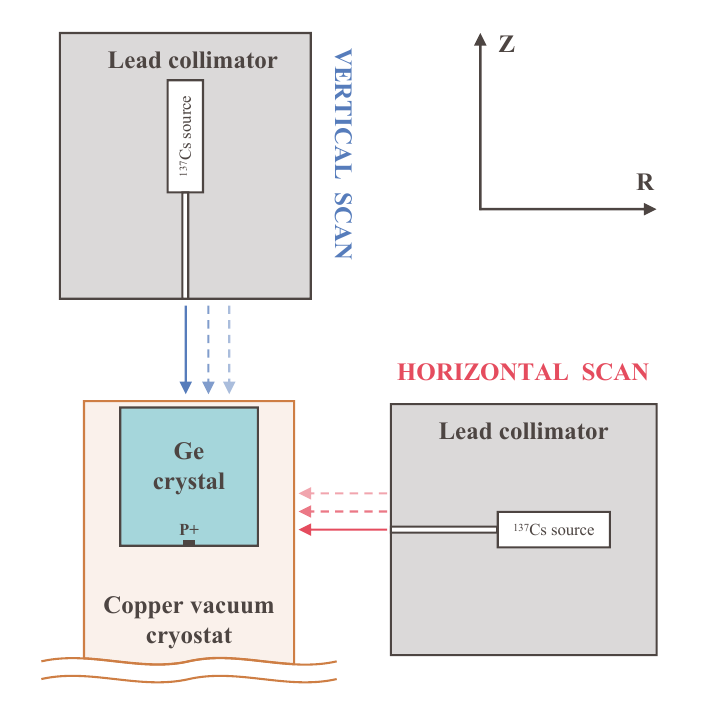}
    \caption{Schematic diagram of the PSCS operational setup, illustrating two scanning configurations: vertical scan and horizontal scan.}
    \label{fig:3}
\end{figure}

The similarity between pulse shapes is quantified using the following $\chi^2$ statistic:
\begin{equation}
    \chi^2 = \frac{1}{N} \sum_{i=0}^{N} (v_i - h_i)^2,
    \label{eq:3}
\end{equation}
where $v_i$ and $h_i$ represent the sampled values of two pulse shapes $v$ and $h$ from datasets $V$ and $H$, respectively, and $N$ denotes the number of sampling points. A smaller $\chi^2$ value indicates higher similarity between the two pulse shapes; theoretically, $\chi^2$ reaches its minimum at the beam intersection. A full-volume scan of the detector can be achieved by translating the collimation direction of the source (e.g., translating the collimator along the axial $Z$ or radial $R$ direction, as illustrated in Fig.~\ref{fig:3}).

Two critical factors influencing the performance of the pulse shape comparison algorithm are the signal-to-noise ratio (SNR) and beam collimation. Due to the relative high energy and the concentration on the collimation line of the photoelectric absorption events within the full-energy peak, only these events are selected for analysis. Furthermore, the selection of the $\chi^2$ threshold for defining pulse shape similarity depends on specific experimental conditions and expected objectives.

Constrained by the specific geometry and single-channel readout characteristics of \textit{p}PCGe detectors, their position sensitivity based on pulse shape discrimination is inferior to that of highly segmented germanium detectors. In particular, considering the structural constraints imposed by the external vacuum cryostat, we can only scan the CDEX-1B detector from top and sides, while the sensitive area for position resolution is located below. Therefore, we propose an optimized PSCS method, the core steps of which are as follows:

\noindent\textbf{Step 1 (self-screening):} The dataset from a single direction is randomly partitioned into two equal subsets (e.g., dividing dataset $H$ into $H_a$ and $H_b$), and a preliminary $\chi^2$ comparison is performed between these two subsets:
\begin{equation}
    \chi_{\text{fir}}^2 = \frac{1}{N} \sum_{i=0}^{N} (h_{ai} - h_{bi})^2.
    \label{eq:4}
\end{equation}
By applying a strict threshold, the events can be effectively confined to the geometric collimation line, while a large fraction of MSEs are pre-rejected because of their poor pulse shape matching degree (i.e., yielding larger $\chi^2$ values). The dataset from the $V$ direction is processed using the identical procedure.

\noindent\textbf{Step 2 (cross-localization):} The filtered remaining datasets from both directions ($V^\prime$ and $H^\prime$) are subjected to a second $\chi^2$ comparison to finally extract the pulse shape response at the geometric intersection.

\subsection{Method validation}\label{sec.3.2}

To validate the effectiveness of this method in characterizing the spatial position responses of the \textit{p}PCGe detector, we establish a comprehensive Monte-Carlo and pulse shape simulation framework in this section. This framework encompasses particle transport and energy deposition simulations based on the Geant4 toolkit \cite{agostinelliGeant4aSimulationToolkit2003}, as well as signal generation simulations based on the Siggen software package \cite{radfordRadforddcIcpc_siggen2025}.

In the Geant4 simulation, precise geometry and material models were constructed for the CDEX-1B detector, its peripheral encasing structures, and the collimator to record the position and energy deposition of each single-hit. Subsequently, the Siggen was employed to calculate the internal physical electric potential, physical electric field, and weighting potential distributions of the detector. Based on this, utilizing the Geant4 output as input, the induced signals for single-hits were generated according to the Shockley-Ramo theorem. In the model, the net impurity concentration of the detector was simplified to a linear gradient profile. Although the initial charge cloud size \cite{boggsAnalyticalFittinggamma2023} and its evolution due to diffusion and self-repulsion effects were incorporated, these factors were found to be negligible in the final results. For each event, the total signal was formed by superposing the signals from single-hits, weighted by their respective energies. To maximize the reproduction of realistic physical pulses, the simulation also accounted for the detector's energy resolution and electronic noise, and involved convolution with the electronics response.

\begin{figure}[!htb]
    \centering
    \includegraphics[width=0.99\linewidth]{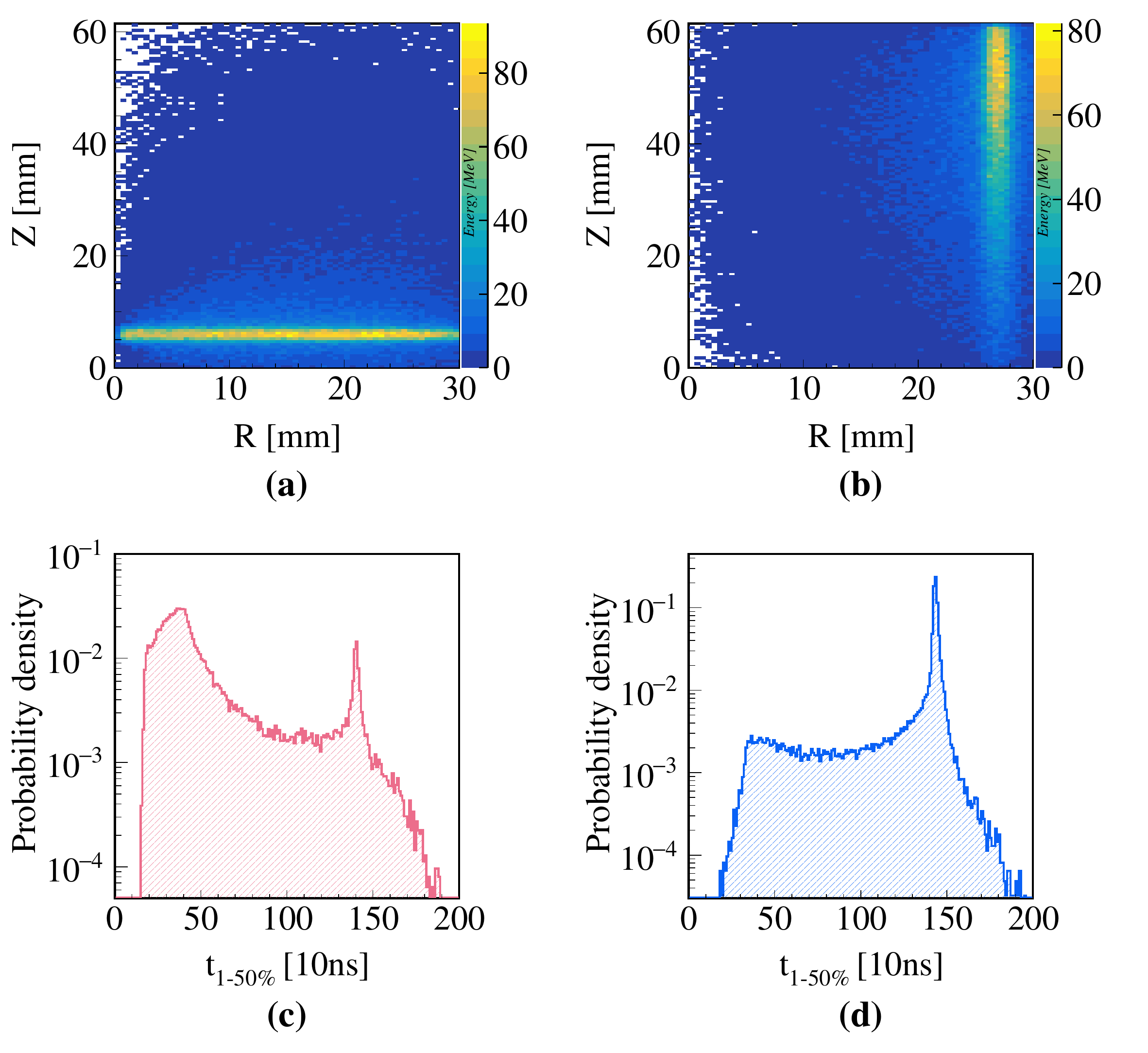}
    \caption{Simulated energy deposition distributions projected on the R-Z plane for (a) the horizontal scan at axial position $Z=6$~mm and (b) the vertical scan at radial position $R=27$~mm. (c) and (d) display the corresponding distributions of the partial rise time $t_{1-50}$ for these two scans, respectively.}
    \label{fig:4}
\end{figure}

Based on the aforementioned framework, we reconstructed the optimized PSCS technical process. The simulation utilized a $^{137}$Cs $\gamma$-ray source (662~keV) collimated by a 1~mm aperture. The scanning procedure consists of two steps: first, a collimated beam is incident laterally, perpendicular to the detector axis at $Z=6$~mm (with the bottom face containing the point-contact defined as $Z=0$~mm), constituting the horizontal dataset ($H$); second, a collimated beam is incident from the top at a radial position of $R=27$~mm, orthogonal to the horizontal beam, constituting the vertical dataset ($V$). Only full-energy peak events were selected for analysis. Fig.~\ref{fig:4}(a) and (b) display the energy deposition distributions of the full-energy peak events in these two datasets, respectively. The results indicate that the vast majority of the energy deposition is accurately concentrated along the prescribed collimation paths.

To quantify the rising characteristics of the pulse shapes, considering the signal properties of \textit{p}PCGe detectors and the electronics settings of CDEX-1B, we define the partial rise time ($t_{1-50}$) as the time interval required for the pulse shape amplitude to rise from 1\% to 50\%. This parameter effectively characterizes the drift time of charge carriers before they reach the vicinity of the point-contact. Fig.~\ref{fig:4}(c) and (d) display the $t_{1-50}$ distributions corresponding to the events in \ref{fig:4}(a) and (b). The distributions exhibit a distinct double-peak structure, corresponding to FBEs along the collimation line, and BEs deviating from the collimation line or MSEs scattered into the bulk region. Since the $t_{1-50}$ values of events within the bulk region are highly consistent, they cluster into a prominent peak in the distribution.

Subsequently, the two datasets were processed using the optimized PSCS algorithm described in Sec.~\ref{sec.3.1}. When calculating the $\chi^2$, the sampling interval primarily focused on the first half of the pulse shape amplitude (i.e., the early rising edge), and segmented $\chi^2$ values were introduced as auxiliary criteria.

\begin{figure}[!htb]
    \centering
    \includegraphics[width=0.99\linewidth]{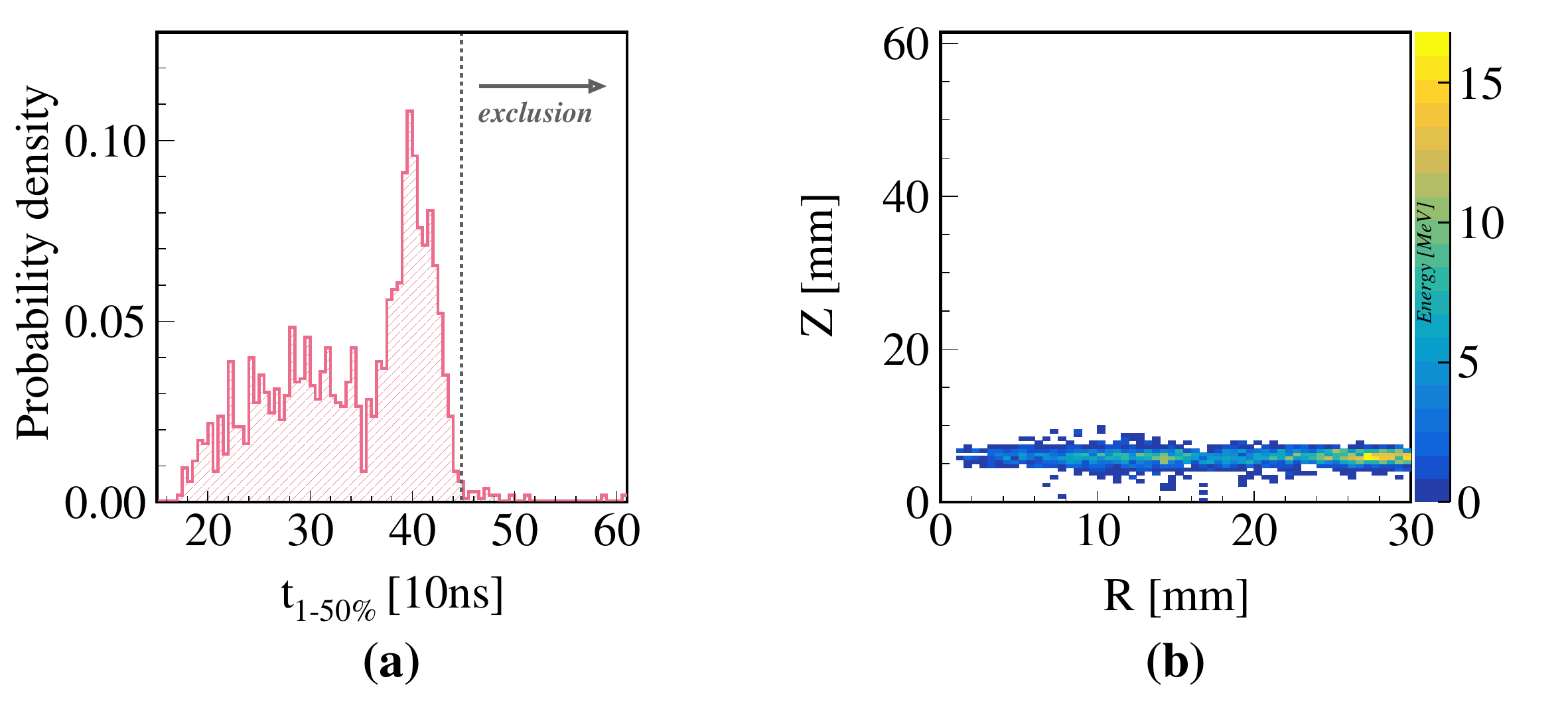}
    \caption{(a) Distribution of $t_{1-50}$ for the remaining events in the horizontal scan dataset after self-screening; the gray dashed line indicates the upper exclusion limit ($+2.5\sigma$) determined by a Gaussian fit to the right-side peak. (b) Energy deposition distribution (R-Z projection) of the remaining events in the horizontal scan.}
    \label{fig:5}
\end{figure}

\noindent\textbf{Step 1: Single-dataset self-screening.} Taking the horizontal dataset $H$ as an example, to accelerate computation, scattered events significantly deviating from the collimation direction are first pre-rejected based on $t_{1-50}$. For the distribution in Fig.~\ref{fig:4}(c), only events with $t_{1-50} < 750$~ns are selected to calculate the $\chi^2$ distribution (Eq.~\ref{eq:4}). This threshold setting is relatively broad (e.g., 600, 800, or 1000~ns), and it does not affect the final screening results. Based on the simulation statistics of this study, the self-screening step retains only the events with the lowest $\chi^2$ values, corresponding to a fraction of 1/30 to 1/15 of the original $H$ dataset. The $t_{1-50}$ distribution of the filtered events is shown in Fig.~\ref{fig:5}(a). Further refinement is performed by Gaussian fitting the main peak and rejecting events beyond the right-side $+2.5\sigma$ limit. Fig.~\ref{fig:5}(b) displays the energy deposition distribution of the remaining dataset $H^\prime$, demonstrating that events are strictly confined to the geometric collimation line and most MSEs have been effectively excluded. The same procedure is applied to the vertical dataset $V$. Notably, the $t_{1-50}$ range determined from $H^\prime$ can be used to constrain the screening of dataset $V$ with extra cross-constraints.

\begin{figure}[!htb]
    \centering
    \includegraphics[width=0.495\linewidth]{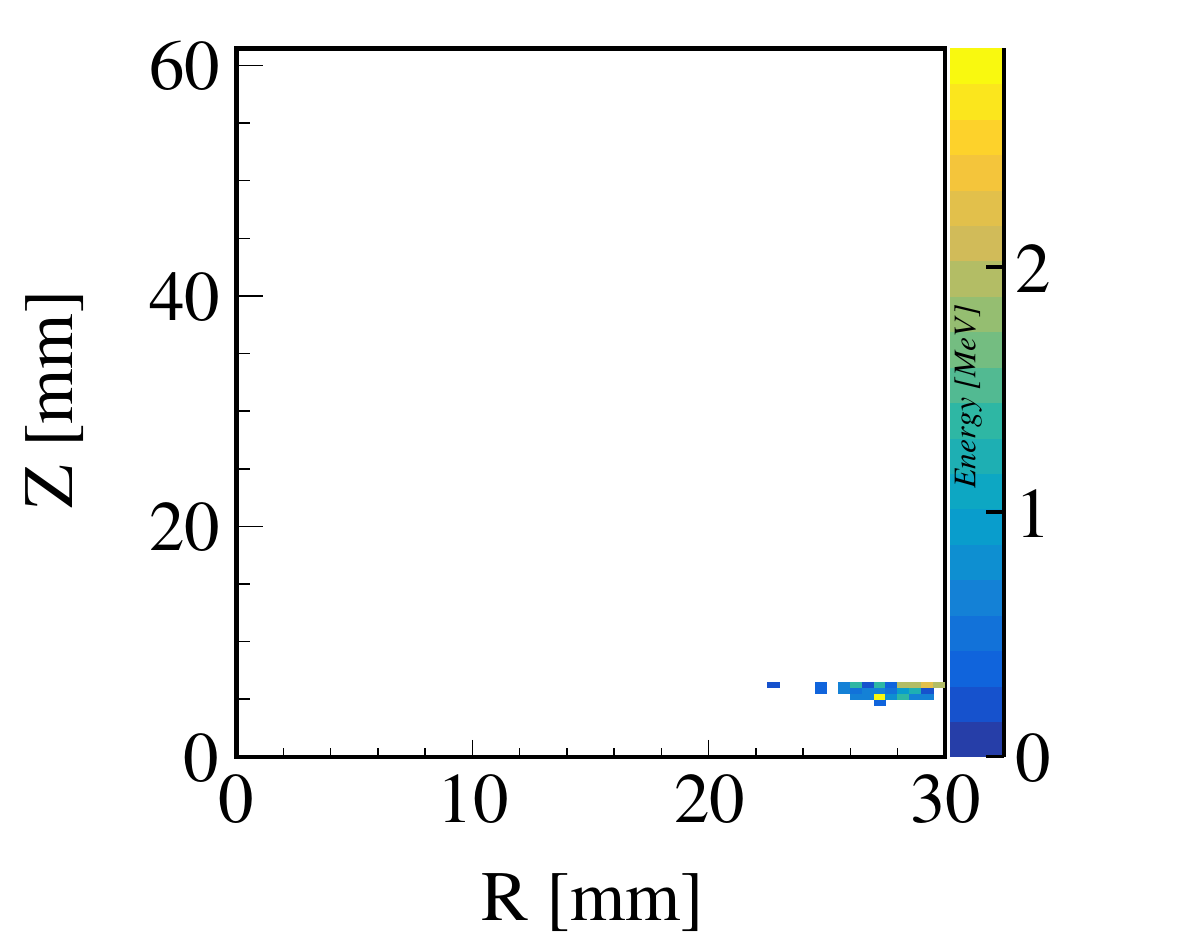}
    \caption{Energy deposition distribution of the final retained events after the second screening algorithm, demonstrating that the events are precisely focused at the expected geometric intersection ($R=27$~mm, $Z=6$~mm).}
    \label{fig:6}
\end{figure}

\begin{figure}[!htb]
    \centering
    \includegraphics[width=0.8\linewidth]{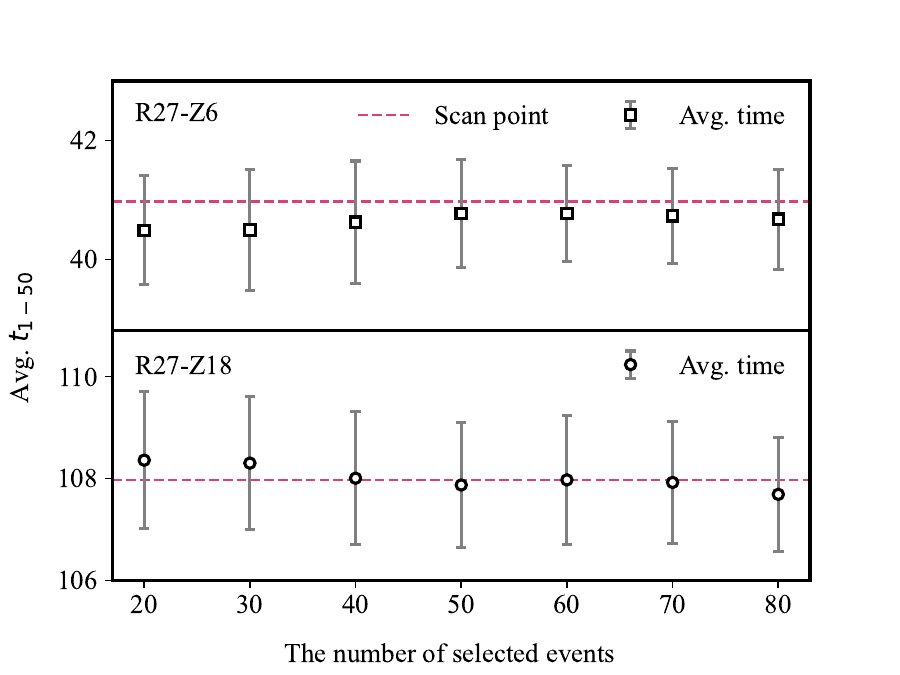}
    \caption{The average $t_{1-50}$ values for two different simulated scanning points (top: R27-Z6, bottom: R27-Z18) when different numbers of events are retained. The red dashed line represents the $t_{1-50}$ value of the pulse shape at the expected scanning position.}
    \label{fig:7}
\end{figure}

\begin{figure}[!htb]
    \centering
    \includegraphics[width=0.75\linewidth]{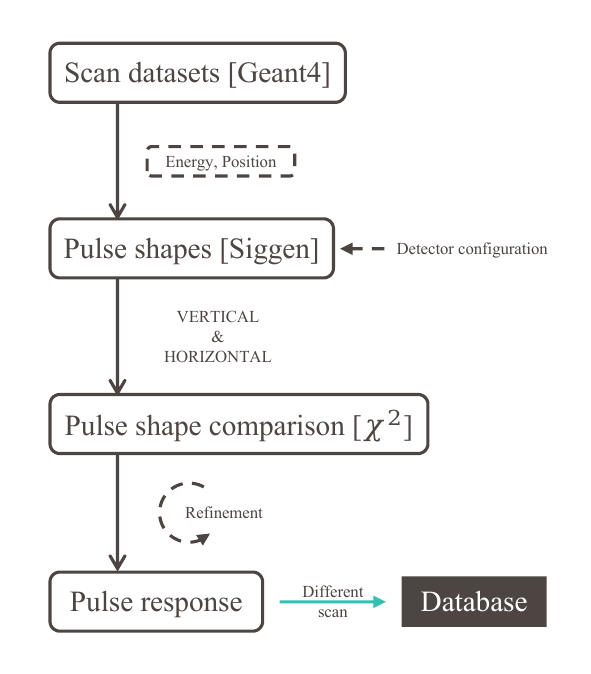}
    \caption{Flowchart of the simulation analysis for the PSCS method.}
    \label{fig:8}
\end{figure}

\noindent\textbf{Step 2: Cross-screening and pulse shape extraction.} Select the filtered remaining datasets $H^\prime$ and $V^\prime$ to calculate the $\chi^2$ distribution (Eq.~\ref{eq:3}), retaining the pairs with the lowest $\chi^2$ values. Considering that the $\chi^2$ algorithm only quantifies the similarity between a specific pair of pulse shapes but cannot guarantee mutual similarity among different pulse pairs, an iterative refinement is required to ensure the consistency of the screening results. This is achieved by calculating the mean $t_{1-50}$ of the retained events, rejecting the 5\% of events deviating furthest from the mean, and repeating this process multiple times \cite{decanditiisSimulationsUsingPulse2020}. Ultimately, 40 to 60 events ($\sim$1/300 of the original dataset) are retained; the distribution of the remaining events is shown in Fig.~\ref{fig:6}. This retained number is an optimal solution derived from a comprehensive trade-off between the pulse rise time characteristics and statistical event fluctuations. Fig.~\ref{fig:7} illustrates the variation of the average $t_{1-50}$ values for two different simulated scanning points when varying numbers of events are retained. The results indicate that keeping the number of retained events between 40 and 60 yields average characteristics that most closely approximate the true pulse shape characteristics at the expected scanning position. Finally, the average pulse shape of these final events is calculated as the standard pulse shape response at that spatial intersection (see the simulation results in Fig.~\ref{fig:10}). Fig.~\ref{fig:8} comprehensively illustrates the simulation flowchart of this method.

\section{Experimental characterization of spatial position responses}\label{sec.4}

\subsection{Experimental setup}\label{sec.4.1}

In this study, a custom-designed scanning system was utilized to measure the position-dependent pulse shape response characteristics of the CDEX-1B detector. The system primarily consists of two functionally independent stainless steel scanning stages and a lead collimator, accommodating the continuous sampling requirements of the PSCS technique across different orientations. The main body of the lead collimator is a cylinder with a length of 180~mm and a diameter of 170~mm. The front section is equipped with a removable collimation insert featuring a 1~mm aperture and a 90~mm depth; the rear section houses a removable cylindrical base with a diameter of 60~mm, inside of which a small source chamber is located for placing the $^{137}$Cs radioactive source. The experiment employed a $^{137}$Cs source with an activity of 15.1~MBq, encapsulated within a small metal capsule. 

During the experiment, the CDEX-1B crystal was first secured by a copper holder with a side thickness of 2~mm and a top thickness of 1~mm; subsequently, the entire assembly was placed at the top of the copper vacuum cryostat in an inverted posture (with the point-contact facing downward) (see Fig.~\ref{fig:3}), leaving a gap of 3~mm between the detector's top surface and the inner wall of the cryostat. The cryostat measures 95~mm in diameter and 228~mm in height, with side and top thicknesses of 2.0~mm and 1.1~mm, respectively, and is cooled via a cold finger connected to a liquid nitrogen dewar. Therefore, during the vertical scan, the collimator was positioned above the cryostat. Under the vertical and horizontal scanning configurations, the distances from the front surface of the collimator to the detector center were 80~mm and 65~mm, respectively. 

The raw signals output from the detector's point-contact were first amplified by a pulsed-reset feedback preamplifier and subsequently fed into a gain-adjustable timing amplifier to maximally preserve the fast rising edge characteristics of the preamplifier signals. Ultimately, the signals were digitized and recorded by a 14-bit flash analog-to-digital converter with a 100~MHz sampling rate. The record length for each signal was 12,000 sampling points, corresponding to a pulse shape time window of 120~$\mu$s. The operating bias voltage of the detector was set to +3750~V. 

\subsection{Distribution of pulse shape responses in different collimated scanning directions}\label{sec.4.2}

This section presents the typical results of the aforementioned scanning experiments, integrated with a comparative analysis of the simulation data. To ensure data purity, for each scanning position, we exclusively selected events with energies within the $\pm 2\sigma$ range of the 662~keV full-energy peak to construct the analysis datasets. Performance tests indicate that the energy resolution of the detector at 662~keV is $\sim$1.3~keV. 

\begin{figure*}[!htb]
    \centering
    \includegraphics[width=0.85\linewidth]{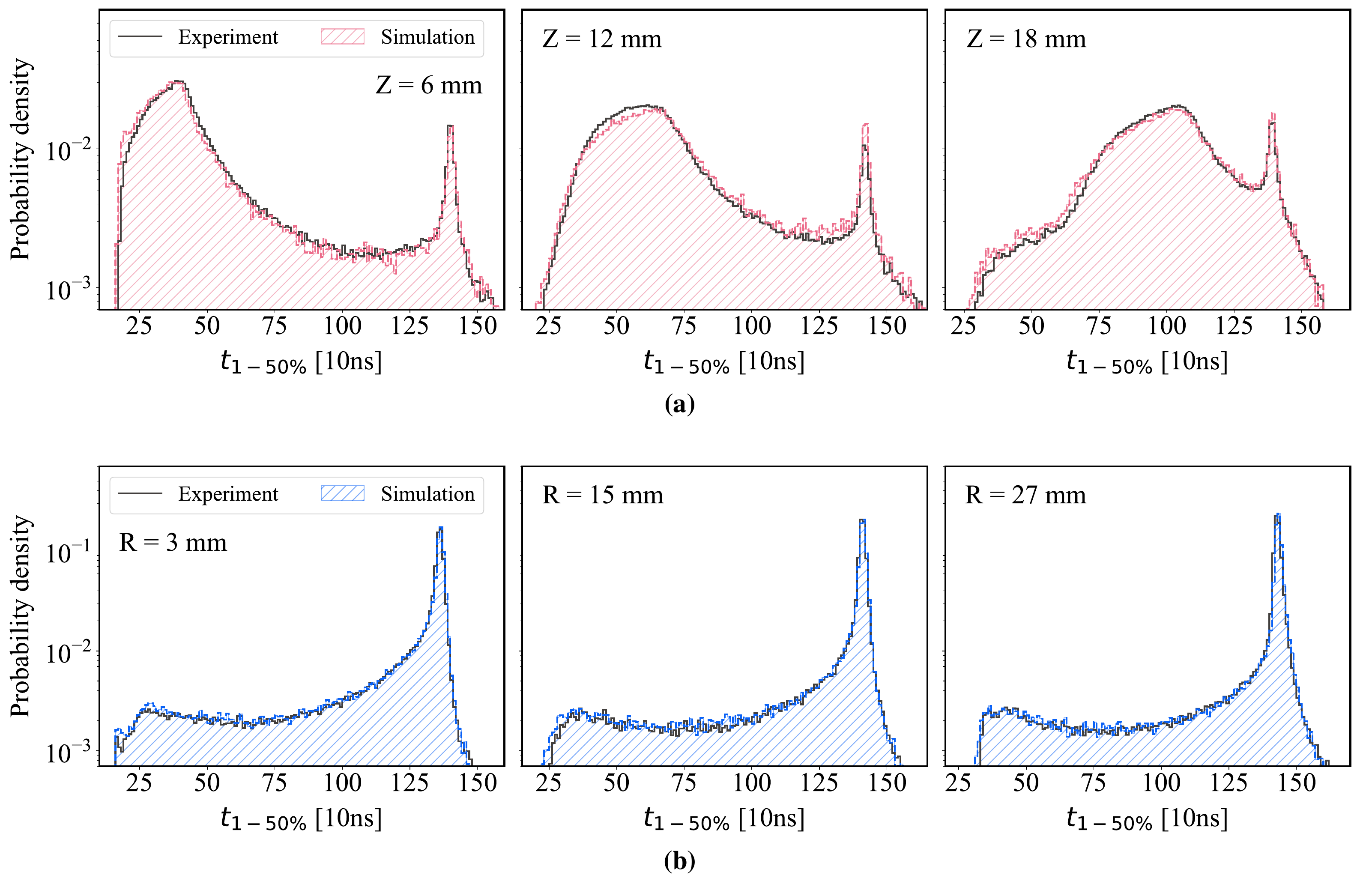}
    \caption{$t_{1-50}$ distribution results for (a) different horizontal scans and (b) different vertical scans. The black lines represent experimental measurements, while the red and blue lines represent the corresponding simulation results.}
    \label{fig:9}
\end{figure*}

To intuitively characterize the pulse shape of events at different scanning positions, we compiled the $t_{1-50}$ distributions for both experimental and simulated pulses, as illustrated in Fig.~\ref{fig:9}. The results demonstrate that for horizontal scans at different $Z$ positions (Fig.~\ref{fig:9}(a)), the $t_{1-50}$ distribution exhibits a pronounced double-peak structure: the left peak corresponds to the FBEs within the focused range of the collimated beam. As the incident $\gamma$-rays penetrate inward, their flux attenuates exponentially within the germanium crystal, causing a certain broadening of this peak, whose peak position is predominantly dictated by events at the outermost periphery of the detector. Notably, as the scanning axial position $Z$ increases, the $t_{1-50}$ value corresponding to this peak exhibits a resolvable increase (i.e., the pulse shape rises more slowly); this directly confirms that the \textit{p}PCGe detector possesses fine single-site spatial position resolution in the fast bulk region; the right peak corresponds to events scattered into the bulk region. As discussed in Sec.~\ref{sec.3.2}, the pulse shapes within this region are highly homogenized, thereby forming a narrow peak. Similar phenomena are also observed in the vertical scans (Fig.~\ref{fig:9}(b)). Because the incident direction of the vertical scan is from top to bottom, the rays traverse a thicker bulk region, resulting in an accumulation of more BEs. Nevertheless, the position-dependent differences exhibited by the FBEs distribution (the left side) remain observable. 

Fig.~\ref{fig:9} demonstrates that the simulation accurately reproduces all the aforementioned key physical phenomena. Regarding the detailed discrepancies between the experiment and simulation, they can be physically attributed to: [1] peak height differences: geometric deviations caused by the irregular distribution and incomplete fixation of the radioactive source's active material in the experiment; [2] local morphological differences: minor deviations between the linear impurity concentration model adopted in the simulation and the actual gradient profile of the crystal. These factors may all influence the initial drift path of carriers and the final pulse shapes. Despite this, the simulation framework has demonstrated extremely high reliability, proving to be fully sufficient for constructing a high-precision pulse shape database and providing a reliable reference baseline for subsequent PSA. 

\subsection{Extraction of pulse shape responses at specific spatial positions}\label{sec.4.3}

\begin{figure*}[!htb]
    \centering
    \includegraphics[width=0.95\linewidth]{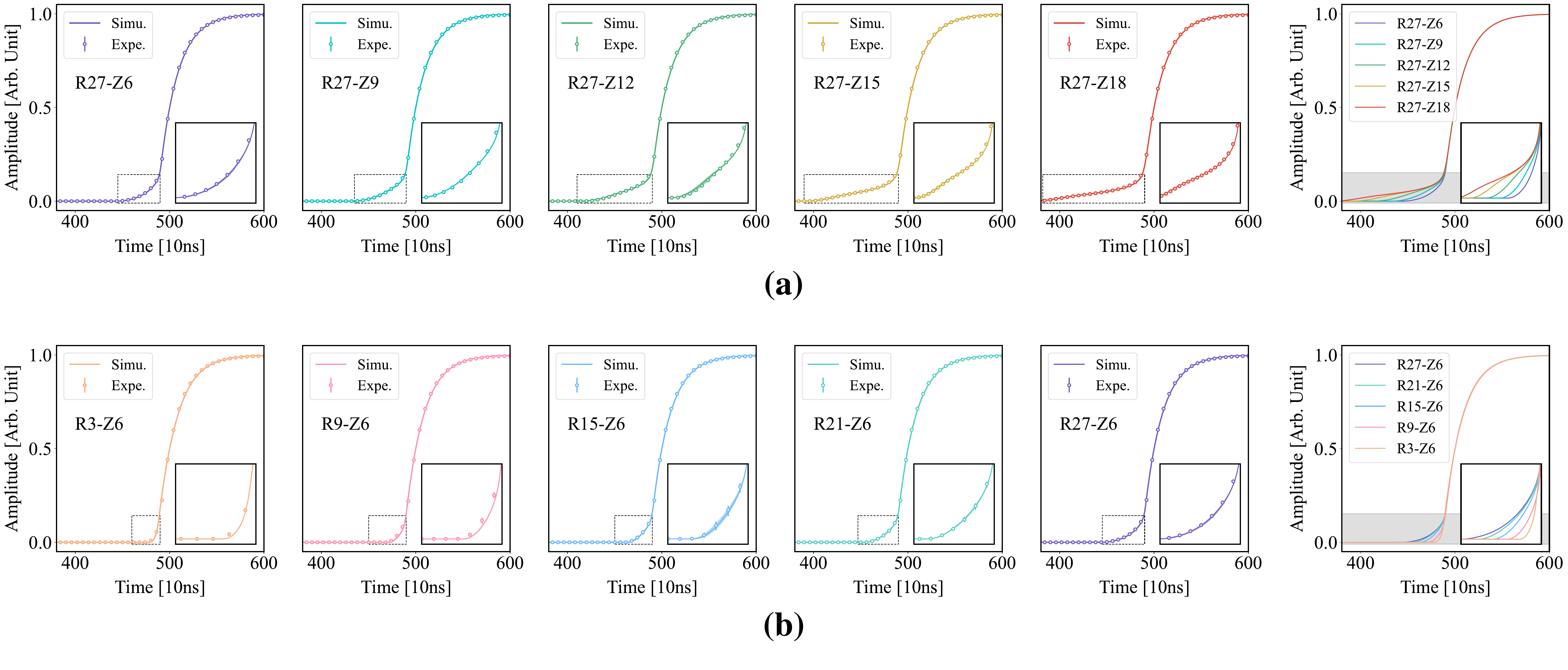}
    \caption{Pulse shape responses at various positions obtained via the PSCS method. (a) Results for different Z positions at a fixed R = 27~mm; (b) Results for different R positions at a fixed Z = 6~mm. Dots denote experiments while curves represent simulations. The fast-rising leading edges of the pulse shapes are highlighted in the insets.}
    \label{fig:10}
\end{figure*}

After normalizing and time-aligning the signals, we extracted the pulse shape responses at the expected positions using the optimized PSCS technique described above. Fig.~\ref{fig:10} illustrates the pulse shape extraction results for typical positions obtained from both experiment and simulation. It is clearly visible that even mm-scale positional changes can lead to significant differences in pulse shape responses. A careful observation reveals an extremely high degree of consistency between the simulated and experimental pulse shapes during the initial rising phase and at the critical kink feature (indicated by the dashed box), which is of paramount importance for event identification based on pulse shapes. Furthermore, the results indicate that the variation of the pulse shape in the $Z$ direction (Fig.~\ref{fig:10}(a)) is significantly greater than that in the $R$ direction (Fig.~\ref{fig:10}(b)), which aligns with the physical characteristics of the weighting potential gradient distribution inside point-contact detectors. 

\section{Intrinsic spatial position resolution}\label{sec.5}

The spatial position resolution of a detector serves not only as a critical metric for evaluating its performance but also fundamentally dictates the upper physical precision limit of PSA, providing essential guidance for the construction of pulse shape databases. To isolate the physical broadening introduced by external factors during the experiment—such as collimator aperture effects and statistical fluctuations in event screening—and to focus exclusively on evaluating the ultimate performance of the detector system itself, this study defines the intrinsic spatial position resolution. This metric characterizes the ultimate capability of the detection system to distinguish the output signal at a single-site position from those in its immediate vicinity; its magnitude is strictly governed by the detector's internal physical signal characteristics and the SNR of the electronics system. Building upon the aforementioned simulation results, this section presents a quantitative investigation into the intrinsic spatial position resolution of the CDEX-1B detector.

\begin{figure}[!htb]
    \centering
    \includegraphics[width=0.8\linewidth]{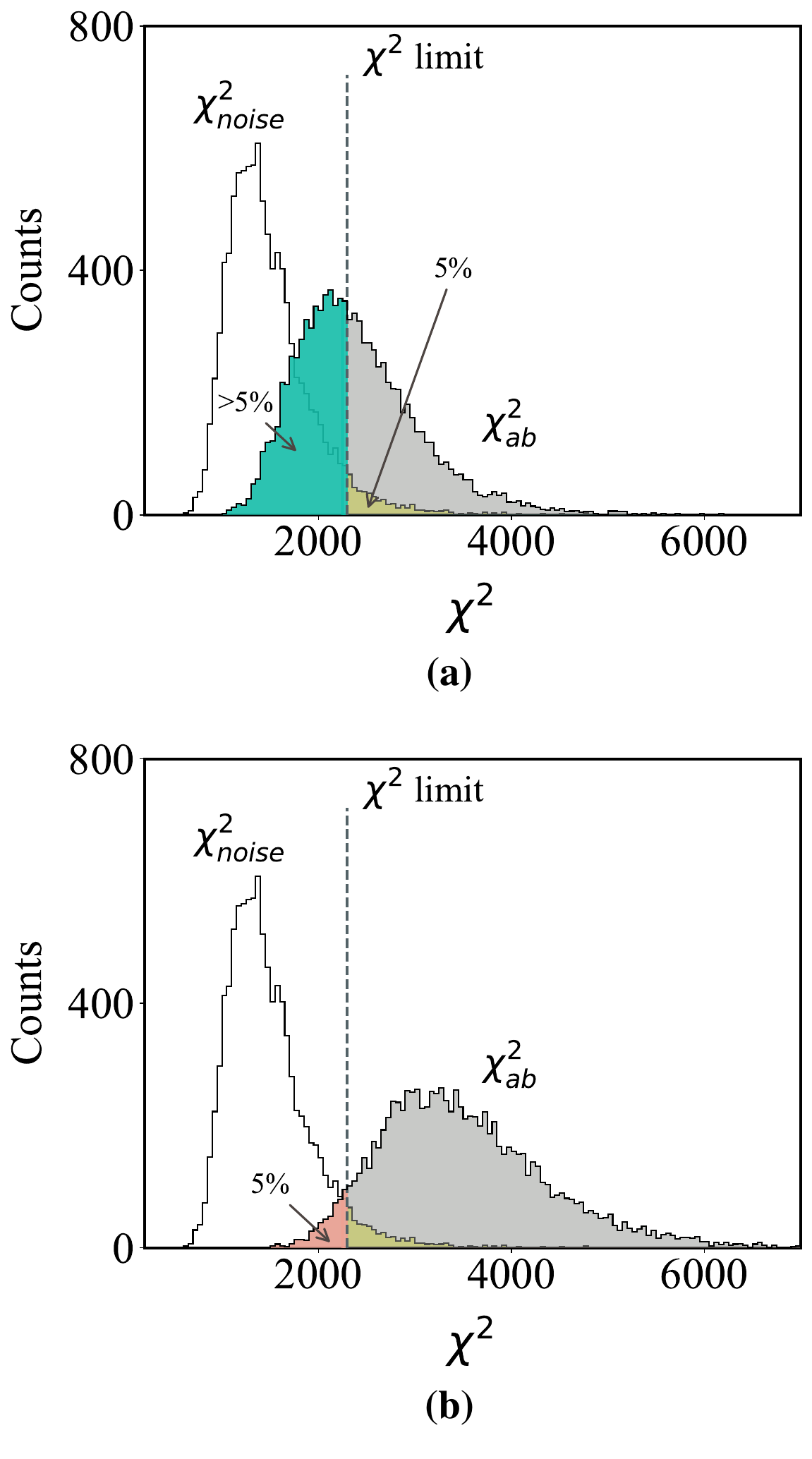}
    \caption{Schematic $\chi^2$ distributions demonstrating the pulse shape discrimination capability. The transparent histograms represent the $\chi_{\text{noise}}^2$ distribution of pulse shapes at the same position affected solely by random noise (self-comparison); the gray histograms represent the $\chi_{ab}^2$ distribution between pulse shapes at position $a$ and an adjacent position $b$ under the influence of noise (mutual-comparison). (a) A scenario where pulse shapes at adjacent positions are indistinguishable (overlap proportion $>$ 5\%). (b) A scenario where pulse shapes at adjacent positions are marginally distinguishable (overlap proportion = 5\%). The vertical dashed line indicates the discrimination threshold ($\chi_{\text{limit}}^2$) defined based on a one-sided 95\% confidence level.}
    \label{fig:11}
\end{figure}

Theoretically, as indicated by Eq.~(\ref{eq:1}) and (\ref{eq:2}), assuming negligible initial charge cloud size and diffusion effects, the induced signals generated at the identical position within the detector should be perfectly consistent following amplitude normalization. Consequently, the experimentally observed discrepancies among pulse shapes originating from the same position stem primarily from electronic noise. The SNR plays a decisive role in this context: a larger deposited energy $q$ yields a smaller relative noise contribution, thereby enhancing the resolution. To quantify the impact of the SNR on pulse shapes at a single position, we construct the following $\chi^2$ statistic:
\begin{equation}
    \chi_{\text{noise}}^2 = \frac{1}{N} \sum_{i=0}^{N} (p_{i,\text{noise1}}^a - p_{i,\text{noise2}}^a)^2,
    \label{eq:5}
\end{equation}
where $p_{i,\text{noise1}}^a$ and $p_{i,\text{noise2}}^a$ denote the sampled values of two normalized pulse shapes at the identical position $a$, superimposed with different random noise sequences ($\text{noise1}, \text{noise2}$), and $N$ represents the number of sampling points. Since the original noise-free pulse shape $p_{\text{ori}}^a$ remains identical, this $\chi^2$ value exclusively reflects the contribution of the system's electronic noise. Fig.~\ref{fig:11} illustrates the self-comparison $\chi_{noise}^2$ distribution of pulse shapes modulated by random noise at the same position, calculated by Eq.~\ref{eq:5}. Furthermore, we evaluate the $\chi^2$ distribution between the pulse shape at position $a$ and that at an adjacent position $b$:
\begin{equation}
    \chi_{ab}^2 = \frac{1}{N} \sum_{i=0}^{N} (p_{i,\text{noise1}}^a - p_{i,\text{noise2}}^b)^2,
    \label{eq:6}
\end{equation}
\noindent the distribution calculated by Eq.~\ref{eq:6} corresponds to the gray histograms in Fig.~\ref{fig:11}.

To quantify the distinguishability between the signal at position $a$ and its adjacent signals, we define the discrimination threshold ($\chi_{\text{limit}}^2$) as the upper limit of the one-sided 95\% confidence interval for the noise-only $\chi_{\text{noise}}^2$ distribution (corresponding to a significance level of $\alpha=0.05$). The criterion is established as follows: if the proportion of samples in the $\chi_{ab}^2$ distribution falling below $\chi_{\text{limit}}^2$ exceeds 5\% (indicated by the green overlapping area in Fig.~\ref{fig:11}(a)), it implies that the pulse shapes at the two positions cannot be effectively distinguished statistically. Conversely, if this proportion is less than 5\% (indicated by the pink area in Fig.~\ref{fig:11}(b)), the pulse shapes at the two positions are considered significantly distinguishable. Based on this criterion, utilizing position $a$ as the starting point and searching along the $R$ or $Z$ direction, the radial or axial distance to the nearest position $b_{\text{min}}$ that satisfies the distinguishability condition is defined as the intrinsic spatial position resolution at position $a$ in that specific direction.

\begin{figure*}[!htb]
    \centering
    \includegraphics[width=0.95\linewidth]{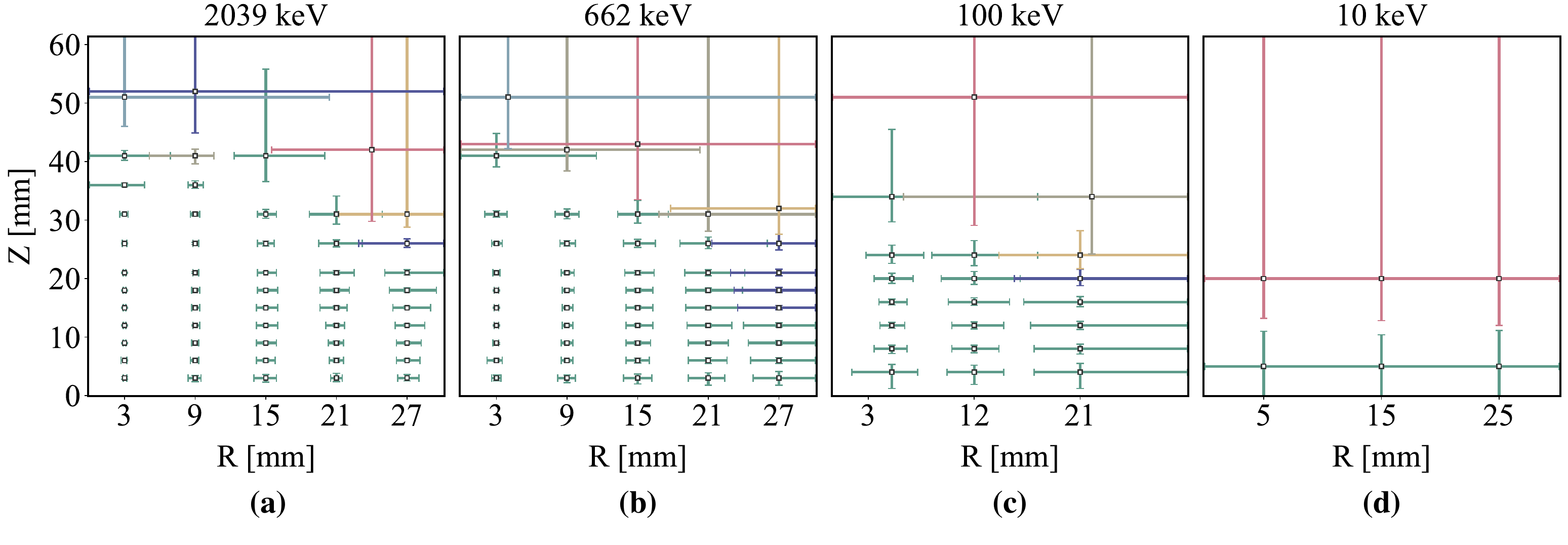}
    \caption{Distributions of the intrinsic spatial position resolution of the CDEX-1B detector under the electronics system of this study for energy regions of (a) 2039~keV, (b) 662~keV, (c) 100~keV, and (d) 10~keV. The lengths of the error bars represent the magnitude of the intrinsic resolution at each grid point along the axial ($Z$) and radial ($R$) directions. Notably, in the highly scrutinized dark matter low-energy detection region ($\sim$10~keV), the detector can only barely distinguish between BEs and FBEs, essentially losing its radial resolution; this conclusion aligns highly with the previous research findings of our group \cite{liIdentificationAnomalousFast2022a}.}
    \label{fig:12}
\end{figure*}

Fig.~\ref{fig:12} presents the simulated intrinsic spatial position resolution distributions of the CDEX-1B detector across multiple critical energy regions, encompassing the dark matter low-energy region ($\sim$10~keV), 100~keV, 662~keV, and the $^{76}$Ge $0\nu\beta\beta$ decay region (2039~keV). This evaluation is strictly based on the electronics system and noise configurations of the aforementioned experiments, while incorporating the impact of crystal axis effects on carrier drift. The results reveal the following:
\begin{itemize}
    \item[\textbf{[1]}] \textbf{Spatial non-uniformity:} Overall, the detector's pulse shape discrimination capability is optimal in the fast bulk region (particularly in the vicinity of the point-contact), and progressively degrades radiating outward from this center.
    \item[\textbf{[2]}] \textbf{Anisotropy:} The resolution in the axial ($Z$) direction is significantly superior to that in the radial ($R$) direction, which strictly maps the physical distribution characteristics of the weighting potential gradient.
    \item[\textbf{[3]}] \textbf{Resolution blind spots:} The bulk region at the top of the detector possesses almost no effective position resolution.
    \item[\textbf{[4]}] \textbf{Energy dependence:} Higher event energies correlate with superior detector resolution.
\end{itemize}
\noindent These conclusions possess fundamental reference value for guiding the future construction of pulse shape databases.

\section{Spatial position tracing of background events and result analysis}\label{sec.6}

\subsection{Construction of the pulse shape database and validation of event localization}\label{sec.6.1}

To meet the stringent requirements of refined PSA regarding spatial grid granularity, this study constructed a simulation-based pulse shape database. The simulation model has been cross-validated with experimental data via the PSCS technique, ensuring the fidelity of its physical responses. Compared to discrete experimental scanning points, the simulation-based database possesses significant advantages, including finer spatial positions, flexible grid generation, and the provision of noise-free standard pulse shapes; furthermore, this database architecture supports modeling corrections by incorporating experimental scanning point data, thereby better accommodating the analysis requirements of actual experimental data. Considering the intrinsic spatial position resolution of the detector determined in Sec.~\ref{sec.5}, this study constructed a non-uniform grid database and primarily applied it to the position reconstruction of the experimental horizontal scanning dataset $H^\prime_{\text{exp}}$ ($Z=6$~mm). This dataset, which predominantly contains SSEs confined near the collimation line, serves as an ideal sample for verifying the localization accuracy of the PSA algorithm.

\begin{figure}[!htb]
    \centering
    \includegraphics[width=0.99\linewidth]{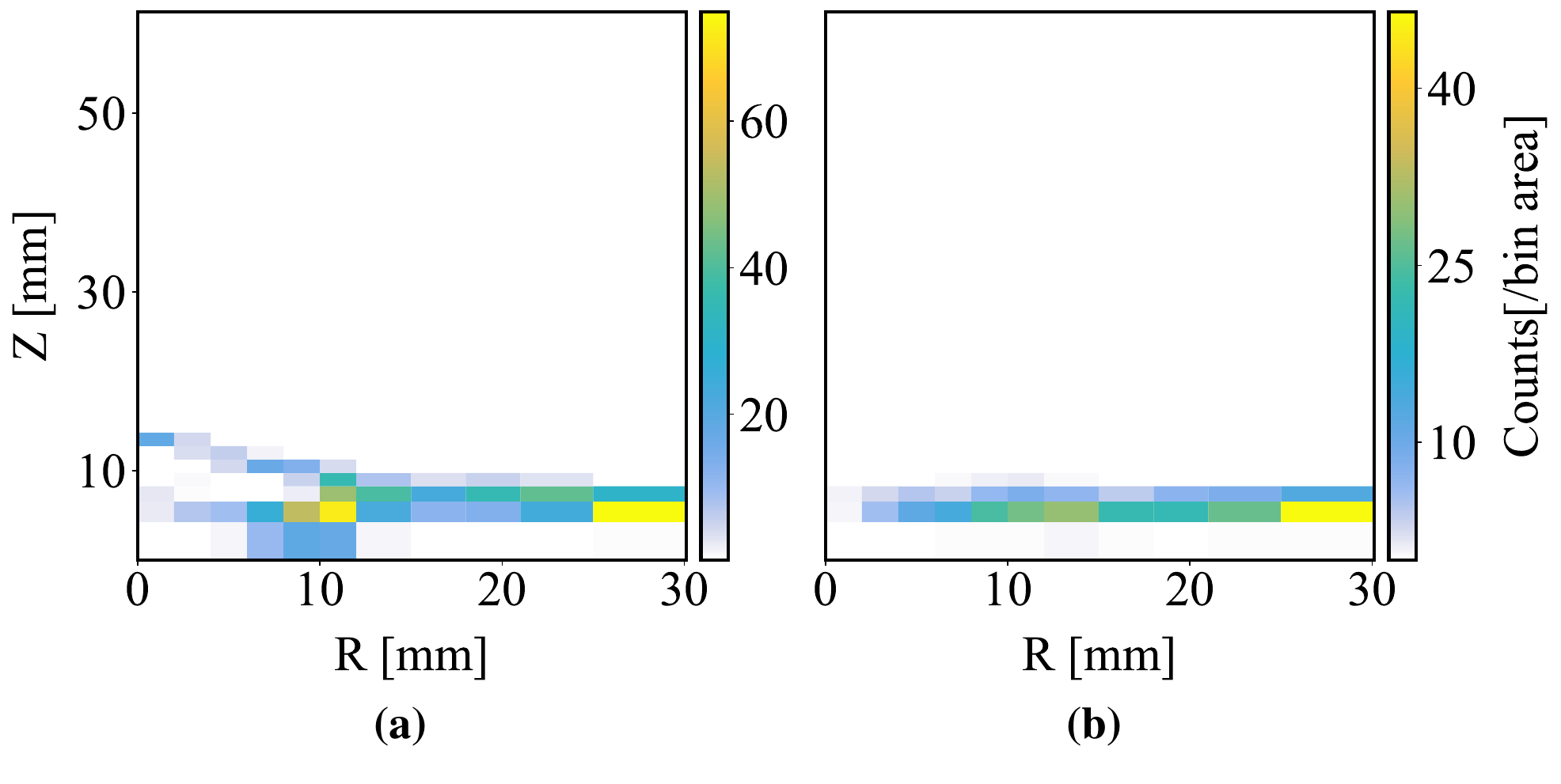}
    \caption{(a) Event position distribution of the experimental horizontal scanning dataset $H^\prime$ ($Z=6$~mm) reconstructed using the PSA technique (event counts are normalized based on grid size). (b) The corresponding expected distribution simulated by Geant4.}
    \label{fig:13}
\end{figure}

Fig.~\ref{fig:13}(a) illustrates the event position distribution reconstructed using the $\chi^2$ matching algorithm. Comparing the experimental reconstruction results with their expected distribution obtained via Geant4 simulations (Fig.~\ref{fig:13}(b)), the two maintain excellent macroscopic morphological consistency: the vast majority of events are accurately focused along the collimation path at $Z=6$~mm, and the radial distribution range aligns with physical expectations. This validates the effectiveness of analyzing real event data using a high-precision simulation-based database.

\subsection{Position tracing of environmental background sources}\label{sec.6.2}

\begin{figure}[htbp]
    \centering
    \includegraphics[width=0.99\linewidth]{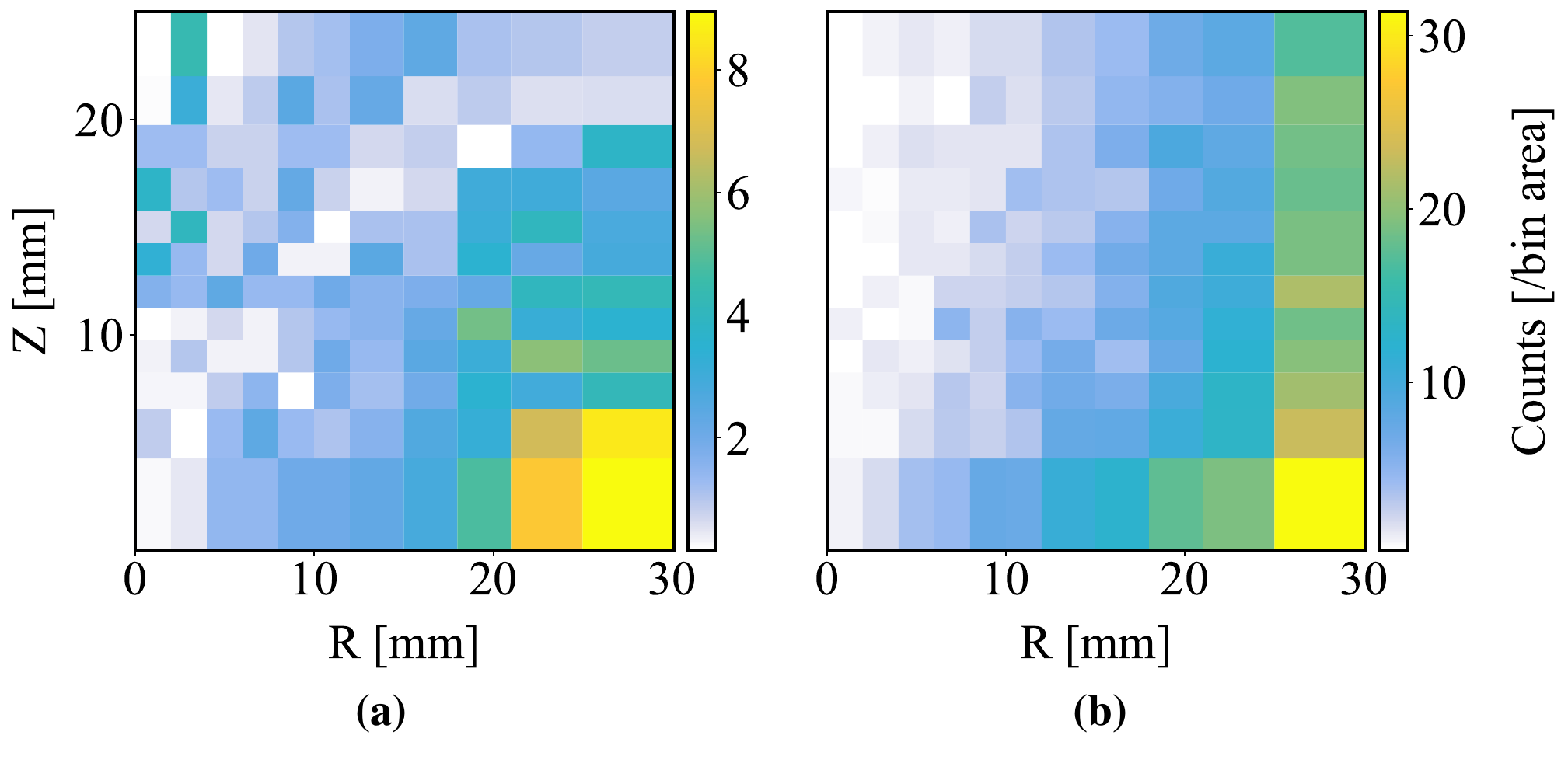}
    \caption{(a) Position reconstruction results of SSEs in the typical Compton continuum within the 610–720~keV energy region from the environmental background during the experiment (only the $Z < 25$~mm region is shown). (b) Expected position deposition distribution of the Compton single-site backgrounds in this energy region under the approximate environmental background model constructed based on Geant4 simulations. The Counts are normalized based on grid size.}
    \label{fig:14}
\end{figure}

Beyond the localization validation utilizing artificial collimated sources, we further applied this method to the background position tracing of the CDEX-1B detector during its experimental commissioning phase under the actual deployment environment at the Sanmen Nuclear Power Plant in Zhejiang. Here, the detector will subsequently serve the RECODE experiment aimed at the precise measurement of CE$\nu$NS \cite{yangRECODEProgramReactor2024}. Fig.~\ref{fig:14}(a) presents the position reconstruction results of SSEs in the 610–720~keV energy region (typical Compton continuum) from the background data. Here, only the region of $Z < 25$~mm, where position resolution is sensitive, is displayed, with counts normalized based on grid precision rather than volume. The physical reconstruction images indicate that these SSEs are significantly accumulated at the bottom of the detector's outer ring. From a radial ($R$) cross-sectional perspective, this accumulation stems from the larger geometric volume of the outer ring region, which consequently deposits more events; conversely, from an axial ($Z$) cross-sectional perspective, the non-uniform distribution of events strongly implies that environmental radiation is more likely to intrude from the bottom of the detector.

From a physical standpoint, these Compton events most likely originate from uranium- and thorium-series radioactive backgrounds present within the experimental apparatus surrounding the detector (such as the side structures or the underlying copper-brick shielding platform). The $\gamma$-rays emitted by these sources undergo multiple scatterings before entering the detector's sensitive volume and depositing energy. To verify this hypothesis, Fig.~\ref{fig:14}(b) presents the expected position deposition distribution of Compton SSEs in this energy region under an approximate environmental background model constructed based on Geant4 simulations. The experimental and simulation results indicate that, although limited by the inability of the preset background model to perfectly replicate the extremely complex actual background environment, leading to minor discrepancies between the two, they maintain a high degree of consistency in their macroscopic distribution features. This powerfully corroborates the aforementioned inference regarding the background sources, providing a crucial physical basis for the efficient suppression of specific backgrounds in the RECODE experiment and future ton-scale detector experiments.

\subsection{The degeneracy of pulse shape characteristics}\label{sec.6.3}

While affirming the localization capability of pulse shape matching, its underlying physical limitations must also be examined. The reconstructed position distribution in Fig.~\ref{fig:13}(a) exhibits a certain "upturn" phenomenon in the region near the detector center ($R < 10$~mm) (i.e., the reconstructed positions shift upwards along the Z-axis), deviating from the expected collimation path. This phenomenon likely originates from the unique physical distribution characteristics of the pulse shape features inside the detector.

\begin{figure}[htbp]
    \centering
    \includegraphics[width=0.6\linewidth]{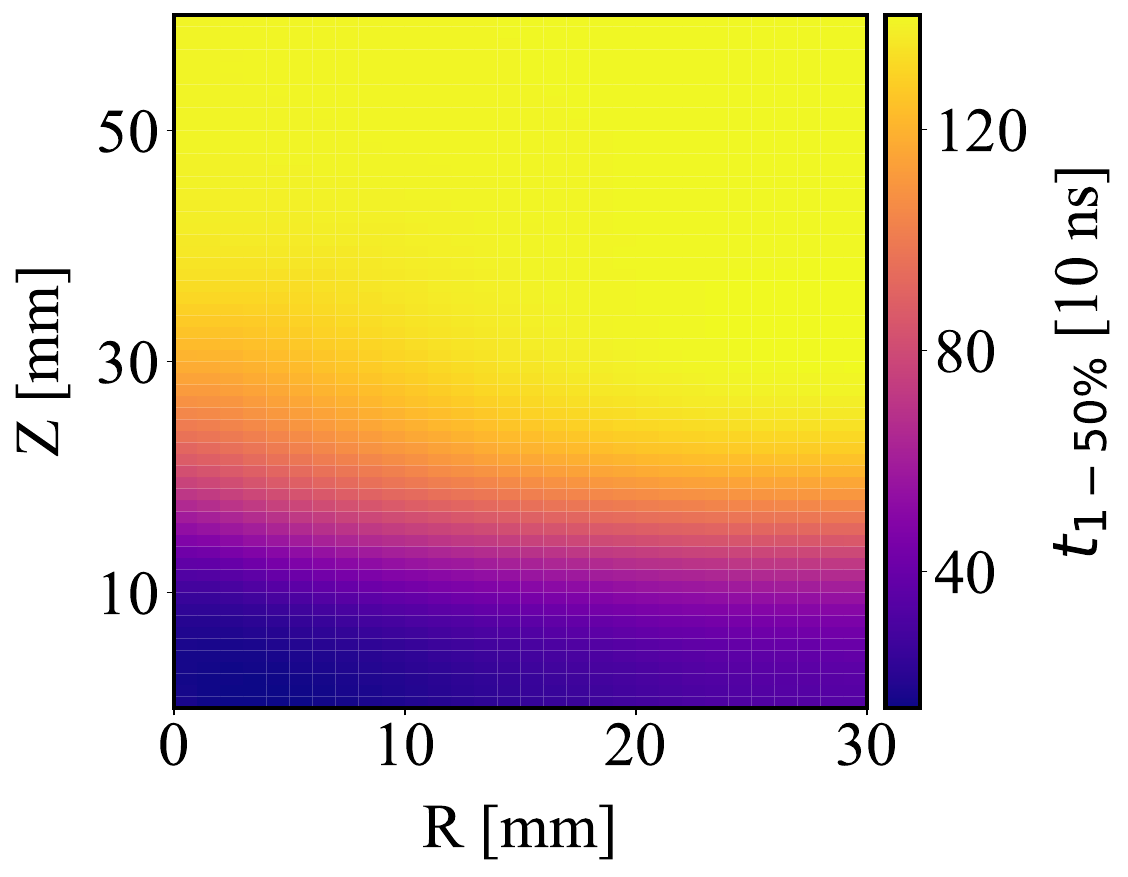}
    \caption{Heatmap of the $t_{1-50}$ distribution for simulated pulse shapes across the entire volume of the CDEX-1B detector. The color scale indicates the magnitude of $t_{1-50}$, intuitively displaying spatial regions with similar rise time characteristics.}
    \label{fig:15}
\end{figure}

To deeply investigate the origin of this deviation, we plotted a heatmap of the $t_{1-50}$ distribution for simulated pulse shapes across the entire detector volume (Fig.~\ref{fig:15}). Observations reveal that the $t_{1-50}$ distribution exhibits significant degeneracy (and its $t_{5-50}$ distribution shares similar characteristics). This implies that, within local regions, the pulse shapes of events located at larger $R$ and smaller $Z$ may be highly similar in their characteristics to those at smaller $R$ and larger $Z$. This homogenized response, induced by the unique weighting potential distribution of the point-contact electrode, renders conventional pulse shape matching algorithms highly susceptible to electronic noise and residual discrepancies between simulation and experiment when processing these regions. It is due to this degeneracy effect that a small fraction of events at the rear end of the collimation path in Fig.~\ref{fig:13}(a) may be misreconstructed to adjacent high-$Z$ positions.

This physical phenomenon profoundly demonstrates that for \textit{p}PCGe detectors, relying solely on a single rise time parameter or the $\chi^2$ comparison of charge signal may be insufficient to fully break this spatial positional ambiguity. In future research, incorporating multi-dimensional pulse shape features or deploying higher-order machine learning frameworks may be the critical pathways to resolving signal degeneracy and further approaching the physical precision limits of PSA.

\section{Conclusion and outlook}\label{sec.7}

Previous studies have demonstrated that \textit{p}PCGe detectors possess unique single-site spatial position resolution potential. Accurately characterizing the position-dependent pulse shape responses inside the detector and evaluating its spatial position resolution are crucial prerequisites for investigating complex background sources and pushing the limits of background suppression in rare-event detection experiments. Focusing on the CDEX-1B detector, this study developed a cross-scanning localization method specifically optimized for the single-channel point-contact geometry, achieving high-precision extraction of pulse shape responses in critical physical regions inside the detector, thereby establishing a solid experimental and theoretical foundation for research on its spatial position resolution.

Relying on the full-chain simulation workflow encompassing Monte-Carlo particle transport and signal generation, this study not only successfully reproduced and validated the experimental scanning process but also, for the first time, quantitatively analyzed the physical limits of the intrinsic spatial position resolution for this type of detector. The evaluation results reveal its non-uniformity and anisotropy, strictly governed by the internal weighting potential gradient distribution: the single-site spatial position resolution of the detector is optimal in the fast bulk region and is significantly superior in the axial ($Z$) direction compared to the radial ($R$) direction; meanwhile, pulse shapes in the vast bulk region exhibit a high degree of homogenization. This physical image elucidates the intrinsic laws of signal evolution and classification in \textit{p}PCGe detectors, providing universal guiding value for spatial position resolution research and pulse shape database construction for similar types of detectors.

Based on the deep integration of the aforementioned simulation framework and real scanning data, this study constructed a non-uniform pulse shape database and ground-breakingly applied it to the position tracing of backgrounds under the actual deployment environment at the Sanmen Nuclear Power Plant. The physical reconstruction images successfully reveal a significant accumulation of Compton SSEs at the bottom of the detector, powerfully corroborating the physical inference regarding the surrounding radiation background sources. Simultaneously, by analyzing the reconstruction deviations in specific regions, it is clarified that the degeneracy of pulse shape characteristics, induced by the point-contact geometry, serves as the deep-seated physical mechanism limiting the accuracy of pulse shape matching and spatial position reconstruction.

In summary, firmly centered on the core physical problem of spatial position resolution, this work has completely established a physical analysis closed-loop for the spatial position response and resolution of \textit{p}PCGe detectors—from the deduction of signal response mechanisms to experimental characterization, and from the quantitative evaluation of intrinsic resolution limits to background tracing in real, complex radiation fields. This not only provides direct scientific evidence for the refined localization and subtraction of complex backgrounds in current experiments but also charts the course for future efforts to incorporate multi-dimensional pulse shape features or higher-order machine learning architectures to break signal degeneracy and approach the physical limits of position resolution. As CDEX advances toward hundred-kilogram and even ton-scale arrays, the spatial position resolution evaluation framework and analysis pathways established in this study will play an extremely critical guiding role in future ultra-low background rare-event search experiments.


\begin{thebibliography}{0}%
\makeatletter
\providecommand \@ifxundefined [1]{%
 \@ifx{#1\undefined}
}%
\providecommand \@ifnum [1]{%
 \ifnum #1\expandafter \@firstoftwo
 \else \expandafter \@secondoftwo
 \fi
}%
\providecommand \@ifx [1]{%
 \ifx #1\expandafter \@firstoftwo
 \else \expandafter \@secondoftwo
 \fi
}%
\providecommand \natexlab [1]{#1}%
\providecommand \enquote  [1]{``#1''}%
\providecommand \bibnamefont  [1]{#1}%
\providecommand \bibfnamefont [1]{#1}%
\providecommand \citenamefont [1]{#1}%
\providecommand \href@noop [0]{\@secondoftwo}%
\providecommand \href [0]{\begingroup \@sanitize@url \@href}%
\providecommand \@href[1]{\@@startlink{#1}\@@href}%
\providecommand \@@href[1]{\endgroup#1\@@endlink}%
\providecommand \@sanitize@url [0]{\catcode `\\12\catcode `\$12\catcode `\&12\catcode `\#12\catcode `\^12\catcode `\_12\catcode `\%12\relax}%
\providecommand \@@startlink[1]{}%
\providecommand \@@endlink[0]{}%
\providecommand \url  [0]{\begingroup\@sanitize@url \@url }%
\providecommand \@url [1]{\endgroup\@href {#1}{\urlprefix }}%
\providecommand \urlprefix  [0]{URL }%
\providecommand \Eprint [0]{\href }%
\providecommand \doibase [0]{https://doi.org/}%
\providecommand \selectlanguage [0]{\@gobble}%
\providecommand \bibinfo  [0]{\@secondoftwo}%
\providecommand \bibfield  [0]{\@secondoftwo}%
\providecommand \translation [1]{[#1]}%
\providecommand \BibitemOpen [0]{}%
\providecommand \bibitemStop [0]{}%
\providecommand \bibitemNoStop [0]{.\EOS\space}%
\providecommand \EOS [0]{\spacefactor3000\relax}%
\providecommand \BibitemShut  [1]{\csname bibitem#1\endcsname}%
\let\auto@bib@innerbib\@empty
\end{thebibliography}%


\begin{thebibliography}{99}


    \bibitem{lukeLowCapacitanceLarge1989a} 
    P. N. Luke, F. S. Goulding, N. W. Madden et al., IEEE Trans. Nucl. Sci., 36(1): 926-930 (1989). \href{http://dx.doi.org/10.1109/23.34577}{doi: 10.1109/23.34577}

    \bibitem{barbeauLargemassUltralowNoise2007} 
    P. S. Barbeau, J. I. Collar, and O. Tench, J. Cosmol. Astropart. Phys., 2007(09): 009 (2007). \href{http://dx.doi.org/10.1088/1475-7516/2007/09/009}{doi: 10.1088/1475-7516/2007/09/009}

    \bibitem{somaCharacterizationPerformanceGermanium2016} 
    A. K. Soma, M. K. Singh, L. Singh et al., Nucl. Instrum. Meth. A, 836: 67-82 (2016). \href{http://dx.doi.org/10.1016/j.nima.2016.08.044}{doi: 10.1016/j.nima.2016.08.044}

    \bibitem{aalsethResultsSearchLightMass2011} 
    C. E. Aalseth et al. (CoGeNT Collaboration), Phys. Rev. Lett., 106(13): 131301 (2011). \href{http://dx.doi.org/10.1103/PhysRevLett.106.131301}{doi: 10.1103/PhysRevLett.106.131301}

    \bibitem{aalsethSearchAnnualModulation2011} 
    C. E. Aalseth et al. (CoGeNT Collaboration), Phys. Rev. Lett., 107(14): 141301 (2011). \href{http://dx.doi.org/10.1103/PhysRevLett.107.141301}{doi: 10.1103/PhysRevLett.107.141301}

    \bibitem{wongSearchNeutrinoMagnetic2007} 
    H. T. Wong, H. B. Li, S. T. Lin et al., Phys. Rev. D, 75(1): 012001 (2007). \href{http://dx.doi.org/10.1103/PhysRevD.75.012001}{doi: 10.1103/PhysRevD.75.012001}

    \bibitem{liLimitsSpinIndependentCouplings2013} 
    H. B. Li, H. Y. Liao, S. T. Lin et al., Phys. Rev. Lett., 110(26): 261301 (2013). \href{http://dx.doi.org/10.1103/PhysRevLett.110.261301}{doi: 10.1103/PhysRevLett.110.261301}

    \bibitem{alvisSearchNeutrinolessDouble2019} 
    S. I. Alvis et al. (Majorana Collaboration), Phys. Rev. C, 100(2): 025501 (2019). \href{http://dx.doi.org/10.1103/PhysRevC.100.025501}{doi: 10.1103/PhysRevC.100.025501}

    \bibitem{arnquistFinalResultMajorana2023} 
    I. J. Arnquist et al. (Majorana Collaboration), Phys. Rev. Lett., 130(6): 062501 (2023). \href{http://dx.doi.org/10.1103/PhysRevLett.130.062501}{doi: 10.1103/PhysRevLett.130.062501}

    \bibitem{acharyaFirstResultsSearch2026} 
    H. Acharya et al. (LEGEND Collaboration), Phys. Rev. Lett., 136(2): 022701 (2026). \href{http://dx.doi.org/10.1103/25tk-nctn}{doi: 10.1103/25tk-nctn}

    \bibitem{legendcollaborationLEGEND1000PreconceptualDesign2021} 
    N. Abgrall et al. (LEGEND Collaboration), arXiv: 2107.11462. \href{http://dx.doi.org/10.48550/ARXIV.2107.11462}{doi: 10.48550/ARXIV.2107.11462}

    \bibitem{cooperPulseShapeAnalysis2011} 
    R. J. Cooper, D. C. Radford, K. Lagergren et al., Nucl. Instrum. Meth. A, 629(1): 303-310 (2011). \href{http://dx.doi.org/10.1016/j.nima.2010.11.029}{doi: 10.1016/j.nima.2010.11.029}

    \bibitem{alvisMultisiteEventDiscrimination2019} 
    S. I. Alvis et al. (Majorana Collaboration), Phys. Rev. C, 99(6): 065501 (2019). \href{http://dx.doi.org/10.1103/PhysRevC.99.065501}{doi: 10.1103/PhysRevC.99.065501}

    \bibitem{cdexcollaborationIntroductionCDEXExperiment2013} 
    K. J. Kang et al. (CDEX Collaboration), Front. Phys., 8(4): 412-437 (2013). \href{http://dx.doi.org/10.1007/s11467-013-0349-1}{doi: 10.1007/s11467-013-0349-1}

    \bibitem{kangStatusProspectsDeep2010} 
    K. J. Kang, J. P. Cheng, Y. H. Chen et al., J. Phys.: Conf. Ser., 203: 012028 (2010). \href{http://dx.doi.org/10.1088/1742-6596/203/1/012028}{doi: 10.1088/1742-6596/203/1/012028}

    \bibitem{chengChinaJinpingUnderground2017} 
    J. P. Cheng, K. J. Kang, J. M. Li et al., Annu. Rev. Nucl. Part. Sci., 67(1): 231-251 (2017). \href{http://dx.doi.org/10.1146/annurev-nucl-102115-044842}{doi: 10.1146/annurev-nucl-102115-044842}

    \bibitem{liuLimitsLightWIMPs2014} 
    S. K. Liu et al. (CDEX Collaboration), Phys. Rev. D, 90(3): 032003 (2014). \href{http://dx.doi.org/10.1103/PhysRevD.90.032003}{doi: 10.1103/PhysRevD.90.032003}

    \bibitem{zhaoSearchLowmassWIMPs2016} 
    W. Zhao et al. (CDEX Collaboration), Phys. Rev. D, 93(9): 092003 (2016). \href{http://dx.doi.org/10.1103/PhysRevD.93.092003}{doi: 10.1103/PhysRevD.93.092003}

    \bibitem{cdexcollaborationLimitsLightWeakly2018} 
    H. Jiang et al. (CDEX Collaboration), Phys. Rev. Lett., 120(24): 241301 (2018). \href{http://dx.doi.org/10.1103/PhysRevLett.120.241301}{doi: 10.1103/PhysRevLett.120.241301}

    \bibitem{maResultsDirectDark2020} 
    H. Ma, Z. She, Z. Liu et al., J. Phys.: Conf. Ser., 1468(1): 012070 (2020). \href{http://dx.doi.org/10.1088/1742-6596/1468/1/012070}{doi: 10.1088/1742-6596/1468/1/012070}

    \bibitem{yangLimitsLightWIMPs2018} 
    L. T. Yang et al. (CDEX Collaboration), Chinese Phys. C, 42(2): 023002 (2018). \href{http://dx.doi.org/10.1088/1674-1137/42/2/023002}{doi: 10.1088/1674-1137/42/2/023002}

    \bibitem{liDifferentiationBulkSurface2014} 
    H. B. Li, L. Singh, M. K. Singh et al., Astroparticle Physics, 56: 1-8 (2014). \href{http://dx.doi.org/10.1016/j.astropartphys.2014.02.005}{doi: 10.1016/j.astropartphys.2014.02.005}

    \bibitem{yangBulkSurfaceEvent2018} 
    L. T. Yang, H. B. Li, H. T. Wong et al., Nucl. Instrum. Meth. A, 886: 13-23 (2018). \href{http://dx.doi.org/10.1016/j.nima.2017.12.078}{doi: 10.1016/j.nima.2017.12.078}

    \bibitem{aguayoCharacteristicsSignalsOriginating2013} 
    E. Aguayo, M. Amman, F. T. Avignone et al., Nucl. Instrum. Meth. A, 701: 176-185 (2013). \href{http://dx.doi.org/10.1016/j.nima.2012.11.004}{doi: 10.1016/j.nima.2012.11.004}

    \bibitem{liIdentificationAnomalousFast2022a} 
    R. M. J. Li, S. K. Liu, S. T. Lin et al., NUCL SCI TECH, 33(5): 57 (2022). \href{http://dx.doi.org/10.1007/s41365-022-01041-x}{doi: 10.1007/s41365-022-01041-x}

    \bibitem{shockleyCurrentsConductorsInduced1938} 
    W. Shockley, Journal of Applied Physics, 9(10): 635-636 (1938). \href{http://dx.doi.org/10.1063/1.1710367}{doi: 10.1063/1.1710367}

    \bibitem{ramoCurrentsInducedElectron1939} 
    S. Ramo, Proc. IRE, 27(9): 584-585 (1939). \href{http://dx.doi.org/10.1109/JRPROC.1939.228757}{doi: 10.1109/JRPROC.1939.228757}

    \bibitem{heReviewShockleyRamo2001} 
    Z. He, Nucl. Instrum. Meth. A, 463(1-2): 250-267 (2001). \href{http://dx.doi.org/10.1016/S0168-9002(01)00223-6}{doi: 10.1016/S0168-9002(01)00223-6}

    \bibitem{crespiNovelTechniqueCharacterization2008a} 
    F. C. L. Crespi, F. Camera, B. Million et al., Nucl. Instrum. Meth. A, 593(3): 440-447 (2008). \href{http://dx.doi.org/10.1016/j.nima.2008.05.057}{doi: 10.1016/j.nima.2008.05.057}

    \bibitem{agostinelliGeant4aSimulationToolkit2003} 
    S. Agostinelli, J. Allison, K. Amako et al., Nucl. Instrum. Meth. A, 506(3): 250-303 (2003). \href{http://dx.doi.org/10.1016/S0168-9002(03)01368-8}{doi: 10.1016/S0168-9002(03)01368-8}

    \bibitem{radfordRadforddcIcpc_siggen2025} 
    D. C. Radford, mjd\_fieldgen and mjd\_siggen software (2025). \href{https://github.com/radforddc/icpc_siggen}{https://github.com/radforddc/icpc\_siggen}

    \bibitem{boggsAnalyticalFittinggamma2023} 
    S. E. Boggs and S. N. Pike, Exp Astron, 56(2-3): 403-420 (2023). \href{http://dx.doi.org/10.1007/s10686-023-09914-8}{doi: 10.1007/s10686-023-09914-8}

    \bibitem{decanditiisSimulationsUsingPulse2020} 
    B. De Canditiis and G. Duchêne, Eur. Phys. J. A, 56(10): 276 (2020). \href{http://dx.doi.org/10.1140/epja/s10050-020-00287-6}{doi: 10.1140/epja/s10050-020-00287-6}
    
    \bibitem{yangRECODEProgramReactor2024} 
    L. T. Yang, Y. F. Liang, and Q. Yue, RECODE Program for Reactor Neutrino CEvNS Detection with PPC Germanium Detector, in Proceedings of XVIII International Conference on Topics in Astroparticle and Underground Physics — PoS(TAUP2023) (University of Vienna: Sissa Medialab, 2024), p. 296. \href{http://dx.doi.org/10.22323/1.441.0296}{doi: 10.22323/1.441.0296}

\end{thebibliography}
\end{document}